\documentclass[lettersize,journal]{IEEEtran}

\usepackage{graphicx}
\usepackage{subcaption}
\usepackage[T1]{fontenc}
\usepackage{amsmath}
\usepackage{amsfonts}
\usepackage{dsfont}
\usepackage{booktabs}
\usepackage{tikz}
\usetikzlibrary{patterns}
\usepackage{todonotes,url}
\usepackage[vlined,linesnumbered]{algorithm2e}
\usepackage{cite}
\usepackage{multirow}
\usepackage{wrapfig}
\usepackage[c]{esvect}
\usepackage{paralist}
\usepackage[font=footnotesize,labelfont=bf]{caption}
\usepackage{xcolor}
\usepackage[normalem]{ulem}
\usepackage{adjustbox}
\usepackage{array}
\usepackage{booktabs}
\usepackage{rotating}
\usepackage{thmtools}
\usepackage{thm-restate}
\usepackage{macros}

\def\BibTeX{{\rm B\kern-.05em{\sc i\kern-.025em b}\kern-.08em
    T\kern-.1667em\lower.7ex\hbox{E}\kern-.125emX}}

\newcommand{\txt}[1]{\textnormal{\normalfont\rmfamily{\textit{#1}}}}
\newcommand{\myvec}[1]{\ensuremath{\vv{#1}}}
\renewcommand{\v}[1]{\ensuremath{\myvec{#1}}}

\ifx\czundefined\final
\newcommand{\changes}[1]{{#1}}
\newcommand{\changescs}[1]{{#1}}

\newcommand{\delete}[1]{\textcolor{red}{\sout{#1}}}
\else
\newcommand{\changes}[1]{#1}

\newcommand{\delete}[1]{}
\fi

\newcommand{\wlogword}[0]{w.l.o.g.\!}

\newcommand{\booleanword}[0]{Boolean}
\newcommand{\subcube}[0]{subcube}
\newcommand{\po}[0]{PO}
\newcommand{\pos}[0]{POs}
\newcommand{\PO}[0]{PO}
\newcommand{\POS}[0]{POs}

\newcolumntype{R}[2]{%
    >{\adjustbox{angle=#1,lap=\width-(#2)}\bgroup}%
    c%
    <{\egroup}%
}
\newcolumntype{P}[2]{%
  >{\begin{turn}{#1}\begin{minipage}{#2}\small\raggedright\hspace{0pt}}l%
  <{\end{minipage}\end{turn}}%
}
\SetKwProg{Fn}{function}{}{}
\SetAlFnt{\small}

\begin{document}

\title{Everything You Always Wanted to Know About \\
Generalization of Proof Obligations in PDR
\thanks{This paper is a slightly extended preprint of the version which appears TCAD: DOI: 10.1109/TCAD.2022.3198260, see https://ieeexplore.ieee.org/document/9855658 for early access.}} %\\ in Bit-Level PDR}

\author{%
    \IEEEauthorblockN{Tobias Seufert\textsuperscript{1}, Felix Winterer\textsuperscript{1}, Christoph Scholl\textsuperscript{1}, Karsten Scheibler\textsuperscript{2}, Tobias Paxian\textsuperscript{1}, Bernd Becker\textsuperscript{1}}\\
    \IEEEauthorblockA{\textsuperscript{1}Institute of Computer Science, Albert-Ludwigs-Universit\"at Freiburg, Germany, \\
          \texttt{\{seufert, winteref, scholl, paxiant, becker\}@informatik.uni-freiburg.de}}\\
    \IEEEauthorblockA{\textsuperscript{2}BTC Embedded Systems AG, Germany, \texttt{scheibler@btc-es.de}}
  }

\maketitle

\begin{abstract}
In this paper we revisit the topic of generalizing proof obligations in bit-level Property Directed Reachability (PDR).
We provide a comprehensive study which
(1) determines the complexity of the problem,
(2) thoroughly analyzes limitations of existing methods,
(3) introduces
    approaches to proof obligation generalization that
    have never been used in the context of PDR,
(4) compares the strengths of different methods from a theoretical point of view,
and
(5) intensively evaluates the methods on various benchmarks from
Hardware Model Checking as well as from AI Planning.
\end{abstract}

\section{Introduction}
%\vspace{-2mm}
\label{sec:introduction}
In 2011, the verification engine PDR resp. IC3 was introduced~\cite{Brad:2011} and is nowadays widely considered as the most powerful algorithm for Hardware Model Checking.
Apart from Hardware Model Checking, PDR is in use on lots of different domains, such as Software Model Checking, Hybrid Systems Model Checking,
or AI Planning~\cite{Cima:2012, Hode:2012, Welp:2013, CGMT:2013, Birg:2014, Komu:2014, Suda:2014, Suen:2020}. It has been lifted to SMT on a wide range of theories. However, in \changes{this work we restrict ourselves to} bit-level PDR.

The idea of PDR is to avoid the
unrolling of the transition relation
as in Bounded Model Checking (BMC)~\cite{Bier:1999}
and to rather replace small numbers of large and hard SAT problems
by many small and easy
%%fw SAT problems
ones
based on a single instance of the transition relation only.
PDR repeatedly strengthens a proof by removing unreachable predecessors of unsafe states.
Thereby, PDR tries to avoid enumerating
single states by putting a lot of effort in generalizing these predecessors to preferably expressive state sets.
\changes{Firstly, PDR generalizes states which are predecessors of the unsafe states, so called \emph{proof obligations} (\pos{}).
\pos{} have already been proven to reach an unsafe state, and should therefore not be reachable from any initial state for the system to be safe.
Furthermore, PDR also generalizes states which are proven to be unreachable from the initial states.
In this paper we
put our focus on the generalization of \pos{} because PDR's efficiency relies heavily on these generalization capabilities~\cite{EMB:2011, Grig:2016} and in contrast to the generalization of unreachable states \cite{Brad:2011,Hass:2013,Ivri:2015}, the generalization of proof obligations did not receive that much attention in research lately}.
Since PDR is in use on such a great variety of applications, these domains pose different challenges to the generalization of \pos{}.
\changes{Exact} methods for the generalization of \pos{} would amount to pre-image computations
requiring quantifier elimination.
Common methods approximate quantifier elimination and build on transition functions instead of general transition
relations.  However, we stress that this is insufficient for a great deal of problem domains.

We discuss exact and approximative generalization techniques for circuits, reverted circuits, circuits with invariant constraints and general transition relations. Our contribution is as follows:
\begin{compactenum}
\item
We show that generalizing \pos{} in PDR is
$\Pi_2^P$-complete
in general. Thus, a non-approximative solution will always have the same complexity as a 2-QBF problem.

\item
We investigate generalization techniques for sequential circuits (\ie{} transition \emph{functions}) which have not been used
in the context of PDR to the best of our knowledge
      and we give a thorough analysis of the detailed reasons why known techniques do not work for general transition relations.

\item
We discuss which methods are applicable to circuits with invariant constraints and which transformations can be used
to enable the correct application of all generalization techniques that are known for circuits.

\item
We introduce methods for the general case of transition relations. This includes approximative as well as exact methods which are based on QBF and MaxQBF solving.

\item
We provide a thorough analysis which methods need which properties of the transition relation to be correct.
In that way we provide users of \po{} generalization methods with a guide telling which methods can be applied in which context.

\item
We provide a thorough comparison of the generalization strengths of the different methods from a theoretical point of view.

\item
From a practical point of view, we present an intensive evaluation of the different methods and also combinations of some of them. We consider HWMCC benchmarks with and without invariant constraints
\cite{Hwmc:2015,Hwmc:2017,Hwmc:2019}, as well as AI Planning benchmarks from the International Planning Competition (IPC).
Our results show that the novel methods can improve on well-established existing generalization methods.
Due to the complementary strengths of the methods the results can be further improved by portfolio applications of
different methods. Even the most expensive methods are able to contribute to
overall runtime
improvements by providing stronger
generalizations. Moreover, exact methods are used to analyze the potential for improvement of approximate methods.
\end{compactenum}

\subsubsection*{Related work}
\label{sec:related}
Since the introduction of PDR \cite{Brad:2011}, there have been several improvements on the efficiency of the original algorithm.
One important insight of \cite{EMB:2011} was the use of a dynamic generalization technique for \pos{} by using ternary simulation (01X simulation) instead of a static cone-of-influence analysis as performed in \cite{Brad:2011}. This greatly affected the algorithm's efficiency \changes{in terms of runtime and memory consumption}. The authors of \cite{Choc:2011} present a similar generalization technique \changes{ -- to which we refer as \emph{lifting} --} which uses a SAT solver call instead of simulation.
\changes{We consider ternary simulation as well as lifting as the two most used standard techniques for the generalization of proof obligations in the context of digital circuits.
We compare them to other in PDR yet unused techniques and even extend the exposition of \cite{Choc:2011} by a detailed analysis of the limitations of lifting.}

In \cite{Birg:2014, Gold:2018}, limitations of these techniques are \changes{also} discussed in the context of spurious \pos{} under abstraction and generalization under invariant constraints. The authors of \cite{SeSc:2018,Seuf:2019} discuss \po{} generalization in the context of Reverse PDR.
\changes{We however give a \emph{general} theoretical analysis of the preconditions for lifting and differentiate between lifting being either incorrect or unable to find generalizations.}

For digital circuits, \cite{Ravi:2004} analyzes two techniques for generalizing counterexamples in BMC.
\changes{The lifting approach which is one of them}
has been adapted to PDR in \cite{Choc:2011}.
\changes{Here we adapt the other one to PDR as well and call it
`Implication Graph Based Generalization' (IGBG) in this paper.
}

In the more general context of PDR on top of SAT modulo theories (SMT), the authors of \cite{Hode:2012} briefly mention a cover approach with don't care values for generalizing \pos{}.
\changes{Technically, this is very much related to the SAT-based cover approach that we discuss for general transition relations.}
Furthermore, there are various works which discuss finding minimal satisfying assignments for CNF formulae \cite{Roor:2006, Fran:2010, Deha:2013, Nime:2014}, which relate to the clause cover approaches that we have incorporated for PDR.

In another field of research, \changes{namely} circuit testing, finding minimal test cubes closely relates to finding minimal \pos{} in PDR. The authors of \cite{Saue:2013, Reim:2014} investigate different techniques including MaxSAT and MaxQBF wrt.~their feasibility and generalization capabilities.
Furthermore, the CEGAR-based approach of \cite{Sche:2016} includes a technique similar
to \cite{Ravi:2004} in order to perform automatic test pattern generation in the
presence of unknown values.

\paragraph*{Structure of the paper}
In Sect.~\ref{sec:preliminaries} we give some preliminaries needed for this
\changes{In Sect.~\ref{sec:complexity} we begin with discussing the complexity of the problem of \po{} generalization in general, we present a wide range of generalization techniques for circuits in Sect.~\ref{sec:circuitmethods} as well as for general transition relations in Sect.~\ref{sse:general}. Finally, we analyze the generalization capabilities of the different techniques from
a theoretical point of view and
introduce further improvements in Sect.~\ref{sec:gencap_impr}.}
An experimental evaluation is given in Sect.~\ref{sec:experiments}, and Sect.~\ref{sec:conclusions} summarizes the results
with directions for future research.

\section{Preliminaries}
\label{sec:preliminaries}
\subsection{Basics and Notations}
\label{sse:basics}
We discuss reachability analysis in finite state transition systems, which
has many applications and can be used, \eg{} for the verification of invariant properties or for finding
plans in AI Planning tasks.

\changes{Finite state transition systems $M = (\{0, 1\}^m,\{0, 1\}^n,I,T)$
describe transitions between states from $\{0, 1\}^m$ under inputs from $\{0, 1\}^n$.
$I \subseteq \{0, 1\}^m$ is the set of initial states, $T \subseteq \{0, 1\}^m \times \{0, 1\}^n \times \{0, 1\}^m$
is the transition relation. There is a transition from state $\v{\sigma} \in \{0, 1\}^m$ to
state $\v{\tau} \in \{0, 1\}^m$ under input
$\v{\iota} \in \{0, 1\}^n$ iff $(\v{\sigma}, \v{\iota}, \v{\tau}) \in T$.
A trace of $M$ is a sequence of states $(\v{\sigma_0}, \v{\sigma_1}, \ldots)$ with
$\v{\sigma_0} \in I$,
$\v{\sigma_j} \in \{0, 1\}^m$  and $\exists \v{\iota_j} \in \{0, 1\}^n$ with
$(\v{\sigma_j}, \v{\iota_j}, \v{\sigma_{j+1}}) \in T$ for all $j \in \mathbb{N}$.
The `reachable states' of $M$ are the states occurring on traces.
The goal of reachability analysis is to either compute all reachable states or to
decide whether some states from a given set are reachable.
For symbolic representations of states, sets and relations
we introduce (present) state variables $\v{s} = (s_1, \ldots, s_m)$,
input variables $\v{i} = (i_1, \ldots, i_n)$, and next state variables
$\v{s}' = (s'_1, \ldots, s'_m)$.
States are obtained by assigning \booleanword{}
values to variables $\v{s}$, inputs by assigning \booleanword{} values to variables $\v{i}$ etc..
The transition relation is then represented by a predicate $T(\v{s},\v{i},\v{s'})$, the
set of initial states of $M$ is identified with a predicate $I(\v{s})$.}
For brevity, we often omit the arguments of the predicates and write them without parenthesis.

\changes{\paragraph*{Hardware Model Checking}
In the context of sequential hardware verification, the transition relation $T$ is derived from a circuit and therefore represents a Boolean \emph{function}
from $\{0, 1\}^m \times \{0, 1\}^n$ to $\{0, 1\}^m$.}
The set of unsafe states (in case of verification of invariant properties)
is represented by a predicate $\neg P(\v{s})$.
\changes{Reachability analysis checks whether
some unsafe state is reachable.}

\changes{\paragraph*{AI Planning}
We consider planning problems which implement the \emph{propositional} STRIPS planning formalism.
A STRIPS planning task $P = (\v{s}, I, G, A)$ is defined by a set of state variables $\v{s}$ with
%again
their next state counterparts $\v{s'}$, a predicate $I(\v{s})$ which identifies the initial states, a predicate $G(\v{s})$ which identifies the goal states, as well as a set of actions $A$.
Encoding schemes like from \cite{Rint:2012} transform planning tasks into reachability problems on finite state transition systems.
The resulting transition relation is not necessarily a function but a \emph{general} transition relation.
Reachability analysis checks whether
some goal state is reachable.
}

A \emph{literal} represents a \booleanword{} variable or its negation.
\emph{Cubes} are conjunctions of literals, \emph{clauses} are disjunctions of literals. The negation of a cube is a clause and vice versa. A \booleanword{} formula in \emph{Conjunctive Normal Form} (CNF) is a conjunction of clauses. As usual, we often represent a clause as a set of literals and a CNF as a set of clauses.
A cube $c = s_{i_1}^{\sigma_{1}} \wedge \ldots \wedge s_{i_k}^{\sigma_{k}}$ of literals over state variables
with $i_j \in \{1, \ldots, m\}$,
$\sigma_{j} \in \{0, 1\}$,  $s_{i_j}^{0} = \neg s_{i_j}$ and $s_{i_j}^{1} = s_{i_j}$
represents the set of all states where $s_{i_j}$ is assigned to $\sigma_{j}$ for all $j = 1, \ldots, k$.
Usually we use letters $c$ or $\hat{c}$ to denote cubes of literals over present state variables,
$d'$ or $\hat{d}'$ to denote cubes of literals over next state variables, and $i$ to denote
cubes of literals over input variables. Sometimes we write
$c(\v{s})$, $d'(\v{s}')$ etc. to emphasize on which variables the corresponding cubes depend.
By \emph{minterms} (often named $m$) we denote cubes containing literals for \emph{all} state variables.
Minterms represent single states.

We assume that the transition relation $T$ of a finite state transition system has been translated into
CNF by standard methods like~\cite{Tse:1968}. Modern SAT solvers~\cite{SS:96} are able to check the satisfiability of \booleanword{}
formulas in CNF.
\changes{We denote a satisfiability check performed by a SAT solver for some formula $F$ by SAT?[$F$].
If the SAT solver terminates, it reports either `satisfiable' or `unsatisfiable'. We use the same terminology for satisfiablity checks of QBF formulas.}

Reachability analysis (\eg{} by PDR) often makes use of special properties of the transition relation $T$.
For instance when $T$ results from a circuit, then it represents a \emph{function}, \ie{} it is
\emph{right-unique} and \emph{left-total}. A relation $T(\v{s}, \v{i}, \v{s}')$ is right-unique iff for all assignments
$\v{\sigma}$ to $\v{s}$ and $\v{\iota}$ to $\v{i}$ there is \emph{at most} one assignment $\v{\tau}$ to
$\v{s}'$ such that $(\v{\sigma}, \v{\iota}, \v{\tau}) \in T$.
$T(\v{s}, \v{i}, \v{s}')$ is left-total iff for all assignments
$\v{\sigma}$ to $\v{s}$ and $\v{\iota}$ to $\v{i}$ there is \emph{at least} one assignment $\v{\tau}$ to
$\v{s}'$ such that $(\v{\sigma}, \v{\iota}, \v{\tau}) \in T$.
Similarly, $T(\v{s}, \v{i}, \v{s}')$ is \emph{left-unique} (\emph{right-total}) iff for all assignments
$\v{\tau}$ to $\v{s}'$ and $\v{\iota}$ to $\v{i}$ there is \emph{at most} (\emph{at least})
one assignment $\v{\sigma}$ to $\v{s}$ such that $(\v{\sigma}, \v{\iota}, \v{\tau}) \in T$.

\subsection{An Overview of PDR}
In this paper we consider Property Directed Reachability (PDR)~\cite{EMB:2011}
(also called IC3\cite{Brad:2011}).

\changes{
Without unrolling the transition relation as in Bounded Model Checking (BMC)~\cite{Bier:1999}, PDR produces sets of clauses for each time step individually with the ultimate goal of finding an inductive strengthening of the safety property $P$ (proof of safety).
We call these sets \emph{time frames} and each time frame $k$ corresponds to a predicate $R_k$ represented by a set of clauses\footnote{
In the following we often identify predicates $R_k$ with the state sets represented by them. We further identify the predicate $T$ with the transition relation it represents.
}.
Hereby, for each time frame $k \geq 1$, PDR proceeds with a new time frame $k+1$ if the clauses created in $R_k$ are sufficient such that $R_k \wedge T \Rightarrow P'$.
Additionally, PDR maintains the invariant that all clauses $\neg c$ from $R_{k+1}$ are \emph{inductive relative} to $R_k$, i.e. $(\neg c \wedge R_k \wedge T) \Rightarrow \neg c'$ for $\neg c \in R_{k+1}$ which is exactly the case if $\neg c \wedge R_k \wedge T \wedge c'$ is unsatisfiable.
As a result, $R_k$ over-approximates the set of states which can be reached from $I$ in up to $k$ steps and thus the state sets represented by the $R_i$ are monotonically increasing in $i$ (for $i \geq 1$).
$R_0$ is always equal to $I$.

\SetKwFunction{BaseCases}{BaseCases}%
\SetKwFunction{Strengthen}{Strengthen}%
\SetKwFunction{Propagate}{Propagate}%
\SetKwFunction{ResolveRecursively}{ResolveRecursively}%
\SetKwFunction{SatGeneralization}{SatGeneralization}%
\SetKwFunction{UnsatGeneralization}{UnsatGeneralization}%
\SetKwFunction{Unsafe}{Unsafe}%
\SetKwFunction{Safe}{Safe}%
\SetKwFunction{strengthened}{strengthened}%
\SetKwFunction{resolved}{resolved}%
\SetKwFunction{propagated}{propagated}%
\begin{algorithm}[t]
\scriptsize
\changes{
\DontPrintSemicolon
\SetKwFunction{Pdr}{Pdr}%
\Fn{\Pdr{$I$, $T$, $P$}}{
	\lIf{\BaseCases{} = `\Unsafe'}{\textbf{return} `\Unsafe'} \label{line:basecases}
	\While{true}{
		\lIf{\Strengthen{} = `\Unsafe'}{\textbf{return} `\Unsafe'}  \label{alg:pdr:line:strengthen}
		$N \gets N + 1$, add new $R_N \gets P$ \tcc*[r]{New time frame.}\label{alg:pdr:line:newframe}
		\lIf{\Propagate{} = `\Safe'}{\textbf{return} `\Safe'} \label{alg:pdr:line:propagate}
	}
}
\caption{PDR: main loop.\label{alg:pdr}}}
\end{algorithm}

\begin{algorithm}[tb]
\scriptsize
\changes{
\DontPrintSemicolon
\Fn{\Strengthen{}}{
	\While(\tcc*[h]{SAT: error pred.}){SAT?[$R_N \wedge T \wedge \neg P'$]} { \label{alg:pdr_strengthen:line:predcheck}
		$m \gets$ satisfying present state assignment \;
		$c \gets$ \SatGeneralization{m}\label{alg:pdr_strengthen:line:satgen} \;
		\lIf{\ResolveRecursively{$c, N$} = `\Unsafe'}{\textbf{return} `\Unsafe'} \label{alg:pdr_strengthen:line:resrec}
	}
	\textbf{return} `\strengthened'  \tcc*[r]{successfully strengthened}
}
\caption{PDR: strengthen the trace.\label{alg:pdr_strengthen}}
}
\end{algorithm}

We present the main loop of PDR in Alg. \ref{alg:pdr}.
In iteration $N$, PDR basically tries to construct error paths of length $N+1$ and starts with
checking whether $R_N \wedge T \Rightarrow P'$ via a SAT solver call with $\txt{SAT?}[R_N \wedge T \wedge \neg P(\v{s}')]$ in \texttt{Strengthen()} (Alg. \ref{alg:pdr_strengthen}).
If the SAT solver reports `satisfiable', a predecessor minterm $m$ is extracted from the satisfying assignment,
$m$ is `generalized' to a cube $c$, and thus $c$ represents
only predecessor states of the
unsafe states. It has to be proven that
there is no path from the initial states to $c$.
To do so, the `proof obligation' (\po{}) $c$ on level $N$ (also called \emph{Counterexample To Induction} (CTI)) has to be recursively resolved.
For \pos{} $d$ on level $k$ in general, \texttt{ResolveRecursively} (Alg. \ref{alg:pdr_resrec}), checks whether the clause $\neg d$ is inductive relative to $R_{k-1}$, i.e. $\neg d \wedge R_{k-1} \wedge T \Rightarrow \neg d'$, leading to new SAT calls
$\txt{SAT?}[\neg d \wedge R_{k-1} \wedge T \wedge d']$ (Alg. \ref{alg:pdr_resrec}, l. \ref{alg:pdr_resrec:line:sat}).

\begin{algorithm}[tb]
\scriptsize
\changes{
\DontPrintSemicolon
\Fn{\ResolveRecursively{$d,k$}} {
	\If(\tcc*[h]{Proof obligation in frame 0.}){$k = 0$} {
		\textbf{return} `\Unsafe' \;\label{alg:pdr_resrec:line:cex}
	}
	\While(\tcc*[h]{pred. in $R_{k-1}$?}){SAT?[$\neg d \wedge R_{k - 1} \wedge T \wedge d'$]}{ \label{alg:pdr_resrec:line:sat}
		$\hat{m} \gets$ satisfying present state assignment \; \label{alg:pdr_resrec:line:minterm}
		$\hat{c} \gets$ \SatGeneralization{$\hat{m}$} \; \label{alg:pdr_resrec:line:satgen}
		\lIf{\ResolveRecursively{$\hat{c}, k - 1$} = `\Unsafe'}{\textbf{return} `\Unsafe'} \label{alg:pdr_resrec:line:newpo}
	}
		$\hat{d} \gets$ \UnsatGeneralization{d}\;\label{alg:pdr_resrec:line:unsatgen}% \tcc*[r]{$d$ unreachable in up to $k$ steps}
		$R_1 \gets R_1 \wedge \neg \hat{d}, \ldots, R_{k} \gets R_{k} \wedge \neg \hat{d}$\;\label{alg:pdr_resrec:line:block}
        \textbf{return} `\resolved'
}
\caption{PDR: recursively resolve \po{} ($d,k$).\label{alg:pdr_resrec}}
}
\end{algorithm}

\begin{algorithm}[t]
\scriptsize
\changes{
\DontPrintSemicolon
\Fn{\Propagate{}}{
	\For {$k \in \{1, \ldots, N - 1\}$, $c$ blocked in $R_k$} {
		\If{$\neg$ SAT?[$R_k \wedge T \wedge c'$]} {
			$R_{k + 1} \gets R_{k + 1} \wedge \neg c$\tcc*[r]{UNSAT: push forward}
		}
		\If{$R_k \equiv R_{k+1}$}{
			\textbf{return} `\Safe' \tcc*[r]{Proof of safety.}\label{alg:pdr_propagate:line:proof}
		}
	}
  \textbf{return} `\propagated'
}
\caption{PDR: propagate blocked cubes forward.\label{alg:pdr_propagate}}
}
\end{algorithm}

If this SAT query is unsatisfiable, then $d$ has no predecessor in
$R_{k-1}$ and therefore $\neg d \wedge R_{k-1} \wedge T \Rightarrow \neg d'$ holds. After a possible generalization into $\hat{d}$ it can be blocked in $R_i$ with $i\in \{1,\ldots,k\}$ by $R_i = R_i \wedge \neg \hat{d}$ (Alg. \ref{alg:pdr_resrec}, l. \ref{alg:pdr_resrec:line:block}).
If the SAT query is satisfiable, a new predecessor minterm $\hat{m}$ has been found
and it is generalized using \texttt{SatGeneralization()} (Alg. \ref{alg:pdr_resrec}, l. \ref{alg:pdr_resrec:line:satgen}) into a \po{} $\hat{c}$ at level $k-1$.

If the strengthening of $R_N$ is sufficient and therefore the SAT call from l. \ref{alg:pdr_strengthen:line:predcheck} in
Alg.~\ref{alg:pdr_strengthen}
is unsatisfiable, we conclude that $R_N \wedge T \Rightarrow P'$, increase $N$ by 1, and continue with the next iteration of PDR in Alg. \ref{alg:pdr} (l. \ref{alg:pdr:line:newframe}) after trying to propagate all blocked cubes forward (l. \ref{alg:pdr:line:propagate}).

The procedure stops, if an error path is found (\po~on level 0) or if
during propagation (Alg. \ref{alg:pdr_propagate}) $R_{k-1}$ and $R_k$ become equivalent (l. \ref{alg:pdr_propagate:line:proof}), \ie{} an inductive invariant
$R_k$ has been found.

The efficiency of the method strongly depends on the success of the mentioned generalizations in \texttt{SatGeneralization()} (l. \ref{alg:pdr_strengthen:line:satgen} of Alg. \ref{alg:pdr_strengthen} and l. \ref{alg:pdr_resrec:line:satgen} of Alg. \ref{alg:pdr_resrec}) and \texttt{UnsatGeneralization()} (l. \ref{alg:pdr_resrec:line:unsatgen} of Alg. \ref{alg:pdr_resrec}).

In this paper we provide a detailed analysis of the generalization of \pos{} in \texttt{SatGeneralization()}.}

Some of the known methods for that purpose assume
special properties of the transition relation $T$ such as the \emph{function} property,
since it results from a digital circuit.
Those properties do not hold in all application contexts, \eg{} they do not hold for transition relations occurring
in AI Planning, for transition relations resulting from circuits with additional invariant constraints, or for transition relations in \emph{Reverse PDR}~\cite{SeSc:2018,Seuf:2019}.

Reverse PDR
computes overapproximations $RR_k$ of
the sets of states from which $\neg P(\v{s})$ can be reached in up to $k$ steps.
As already observed in~\cite{EMB:2011} and~\cite{Hass:2013}, there is a simple way to
arrive at an implementation of Reverse PDR based on the fact that
there is a path from
$I(\v{s})$ to $\neg P(\v{s})$ using a transition relation $T$ iff there
is a path from $\neg P(\v{s})$ to $I(\v{s})$ using the `reverted transition relation'.
Thus, a basic version of Reverse PDR is obtained just by exchanging $I(\v{s})$ with $\neg P(\v{s})$
and interpreting the predicate for $T$ `the other way around'.

\changes{\section{Generalization of \POS{} and Its Complexity}\label{sec:complexity}
Generalization plays a crucial role for the efficiency of PDR~\cite{EMB:2011, Grig:2016}.
As explained above, generalization in PDR takes place, when clauses are learnt as well as when new \pos{} are created.
In this paper we restrict our attention to the latter type of generalization.}

Assume that we try to resolve a \po{} $d$ by a call
$\txt{SAT?}[\neg d \wedge R_{i-1} \wedge T \wedge d']$, but
the SAT solver returns a (full) satisfying assignment
with a minterm $m$ representing a single current state.
$m$ is then a new \po{}, but before trying to resolve this \po{}
we try to generalize it into a shorter cube $c$.
The question, whether a given \subcube{} $c$ of $m$ is still a \po{} with successors in $d$,
can be formulated as the following problem:
\begin{definition}
(PO Generalization Problem (POGP))
Given a transition relation $T(\v{s},$ $\v{i},\v{s}')$, a cube
$c = s_{1}^{\sigma_{1}} \wedge \ldots \wedge s_{k}^{\sigma_{k}}$
over present state variables, and a cube $d' = (s_{1}')^{\tau_{1}} \wedge \ldots \wedge (s'_{l})^{\tau_{l}}$
over next state variables\footnote{
To simplify notations we assume here \emph{\wlogword{}} that the variables not occurring in $c$ and $d'$ are
at the end of the vector of state variables.},
decide whether \emph{for all} $(\sigma_{k+1}, \ldots, \sigma_m) \in \{0, 1 \}^{m-k}$ \emph{there is}
$(\tau_{l+1}, \ldots, \tau_m) \in \{0, 1 \}^{m-l}$ and an input $\v{\iota} \in \{0, 1\}^n$, such
that $T(\sigma_1, \ldots, \sigma_m, \v{\iota}, \tau_1, \ldots, \tau_m) = 1$, \ie{}
such that there is a transition from $(\sigma_1, \ldots, \sigma_m)$ to $(\tau_1, \ldots, \tau_m)$ under
input $\v{\iota}$.
\end{definition}\label{def:lifting}

The problem formulation contains a quantifier alternation which is already an indicator for the hardness of
POGP.
\begin{theorem}
\label{pogpproof}
POGP
is $\Pi_2^P$-complete.
\end{theorem}

\begin{IEEEproof}
We show that POGP is $\Pi_2^P$-hard by reducing 2-QBF to POGP.
We consider a 2-QBF formula $\phi = \forall \v{x} \exists \v{y} : \Phi(\v{x}, \v{y})$
with $\v{x} = (x_1, \ldots, x_p)$, $\v{y} = (y_1, \ldots, y_n)$.
Now define $T(\v{s}, \v{i}, \v{s}') := s_1 \wedge \Phi(\v{x}, \v{y}) \wedge s_1' \wedge \ldots \wedge s_{p+1}'$
with $\v{s} := (s_1, x_1, \ldots, x_p)$, $\v{i} = \v{y}$, $\v{s}' = (s_1', \ldots, s_{p+1}')$.
Further define $c = s_1$ and $d' = s_1' \wedge \ldots \wedge s_{p+1}'$.
The defined instance of POGP asks whether for all $\sigma_2, \ldots, \sigma_{p+1} \in \{0, 1\}^p$
there is $\v{\iota} \in \{0, 1\}^n$ such that
$T(1, \sigma_2, \ldots, \sigma_{p+1}, \v{\iota}, 1, \ldots, 1) = \Phi(\sigma_2, \ldots,$ $\sigma_{p+1}, \v{\iota}) = 1$.
The answer is yes iff $\forall \v{x} \exists \v{y} :  \Phi(\v{x}, \v{y})$ is satisfiable.

POGP is \emph{in} $\Pi_2^P$, since its answer is yes iff the 2-QBF
$
\forall s_{k+1} \ldots \forall s_m \exists i_1 \ldots \exists i_n \exists s'_{l+1} \ldots \exists s'_{m} :
c(\v{s}) \wedge T(\v{s}, \v{i}, \v{s}') \wedge d'(\v{s}')
$
is satisfiable.
\end{IEEEproof}

\changescs{
The proof of Theorem~\ref{pogpproof} shows that POGP can basically be viewed as a 2-QBF problem.
From a different point of view, POGP asks whether the cube $c$ is
an
implicant of the
\booleanword{} function $\Phi(\v{s}) := \exists \v{i} \exists \v{s}' : T(\v{s}, \v{i}, \v{s}') \wedge d'(\v{s}')$.
This point of view does not change the complexity of the problem and to take advantage of this view
algorithmically, we would have to perform symbolic elimination of the quantifiers
$\exists \v{i}$ and $\exists \v{s}'$ before considering implicants (or \emph{prime} implicants to make the cube $c$
as short as possible).
}

Due to the high complexity of the problem we first look into approximate solutions in the
\changes{next two sections.}
We start \changes{in Sect.~\ref{sec:circuitmethods}} with the special case of sequential circuits and continue with the general case \changes{in Sect.~\ref{sse:general}}. For the general case we
consider an exact method as well. \changes{In Sect.~\ref{sec:gencap_impr} we compare the strengths of the different methods,
analyze the effectiveness of approximate solutions
for the special case of left-unique transition relations (motivated by Reverse PDR), and finally discuss further improvements.}

\changes{\section{Approximative PO Generalization for Circuits}\label{sec:circuitmethods}}
For the special case of digital circuits, where the transition relation represents a \emph{function},
different approximations of POGP can be used.

\changes{Firstly, we give a short overview of the commonly used techniques for circuits in Sect. \ref{sse:standard}.
These} are 01X-simulation as proposed in~\cite{EMB:2011} as well as the lifting approach proposed in~\cite{Choc:2011}
which is based on a technique of lifting BMC counterexamples from~\cite{Ravi:2004}.
\changes{Besides 01X-simulation and lifting, we also consider the justification technique which is  implemented as an optional PO generalization technique in ABC's \cite{Bray:2010} PDR implementation a known `standard method'.}

While it is rather obvious that 01X-simulation of circuits does not apply for general transition relations, it is more subtle in the case of the lifting approach.
\changes{Therefore we thoroughly discuss the limitations of lifting and additionally discuss extensions which may improve its generalization capabilities in Sect. \ref{sse:lifting_ext}.}
We also present two techniques for circuits which have not been used in the context of PDR yet (to the best of our knowledge): (1) It is feasible to find a state with the maximum amount of X-valued state-bits by using a 01X-encoding of the circuit and a MaxSAT solver (similar to~\cite{Saue:2013}\changes{, see Sect.~\ref{sse:ms01x}}).
Note that \cite{Roor:2006} and \cite{Fran:2010} discuss an approximate version of this method
which uses a SAT solver with an appropriate decision heuristics \changes{(see Sect.~\ref{sse:ternsat})}.
(2) Additionally, for BMC, \cite{Ravi:2004} presents an alternative to the mentioned lifting technique which is based on a reverse traversal of the implication graph of a SAT solver.
This method -- \changes{we call it IGBG} -- can be adapted to the PDR case, too \changes{(see Sect.~\ref{sec:impl})}.

Finally, at the end of this section, we consider the case of circuits with invariant constraints which
lead to transition relations not representing functions \changes{(see Sect.~\ref{sec:selfloops})}.
\changes{\subsection{Standard Methods}\label{sse:standard}}
\subsubsection{01X-Simulation}
\label{01X}
This approach uses a three-valued logic with a don't care value $X$ (the two-valued semantic can be extended by $(X \wedge 0 = 0), (X \wedge 1 = X), (X \wedge X = X), (\neg X = X)$).
We start with a (full) satisfying assignment to $\neg d \wedge R_{k-1} \wedge T(\v{s}, \v{i}, \v{s}') \wedge d'(\v{s}')$ leading to a
\po{} state $m$.
Now, present state bits from $m$
are iteratively assigned to $X$ followed by a simulation of the circuit. If an $X$ propagates to an output which is asserted
by $d'(\v{s}')$, the state bit is necessary in $m$, otherwise it is redundant and can be removed from $m$.
The process is greedily iterated until no more redundant state bits are found.
Apart from the greedy search for redundant state bits and from the fact that only predecessors of $d'$
under a fixed input assignment $i$ are considered, ternary simulation has an additional source of non-optimality:
As an example consider
an AND-gate with output $b$ where both inputs are just the negation of each other: $b \leftrightarrow (a \wedge \neg a)$. If $a$ is assigned to $X$, the $X$ will propagate to $b$ using the rules of three-valued logic even though $b$ is constant--0.
Since this method uses ternary simulation of circuits, it is inherently restricted to transition relations resulting from
circuits (which are transition \emph{functions}).

\subsubsection{Justification Based Generalization\label{sec:justi}}
\changes{A technique strongly related to 01X-simulation is to apply justification to the circuit.}
Given a full assignment \changescs{$m$ to all present state variables, $i$ to all primary inputs,
and $d'$ to a subset of the next state variables}, we
\changescs{look for}
a \emph{partial} assignment to
\changescs{the present state variables}
which is still able to justify resp. imply the assignment $d'$.

In principle, we traverse the circuit and \changescs{heuristically} determine the variables of $m$ which are (together with all variables from $i$) sufficient to imply $d'$.

\changescs{
First of all, the circuit is simulated with the assignment $m$ and $i$.
Second, the literals of all present state variables which are not included in the
syntactical support set of the next state variables contained in $d'$ are
removed from $m$.
Then, priorities are assigned to all primary input variables ($\infty$) and to all
variables of the remaining literals in $m$ (arbitrary natural numbers).
We prefer to keep variables with a higher priority in the assignment.
All input literals receive priority $\infty$, because $i$ remains untouched and therefore, if a next state assignment can be justified by an input or a state variable, we will always prefer to use the input and ignore the state variable.
Now an iterative procedure is started.
In a first iteration, the priorities are forward propagated from the present state variables and primary inputs
towards the next state variables (outputs of the circuit).
The priority of a (circuit input or gate output) variable $v$ is denoted by
$\mathit{prio}(v)$ in the following.
By this propagation, the method implicitly constructs justification paths from the circuit inputs
to the next state variables in $d'$.
Consider a gate with output $z$ and inputs $x$, $y$.
If the value at $z$ is only justified by $x$ and not by $y$, then
$\mathit{prio}(x)$ is propagated to $z$, since there is no choice for justification.
If the value at $z$ can be justified by $x$ \emph{or} by $y$, then the higher
priority is propagated to $z$, since we prefer justification paths starting from variables with high priority.
If both values of $x$ \emph{and} $y$ are needed to justify the value at $z$, then the lower
priority is propagated to $z$ in order to remember overall the input with the lowest priority which
is connected to $z$ by a justification path.
For an AND-gate $z \leftrightarrow x \wedge y$, e.g., this leads to the following rules:
\begin{enumerate}
\item If $z = 0$ and $x \oplus y = 1$, then $\mathit{prio}(z) = \mathit{prio}(\mathit{min}(x,y))$.
\item If $z = 0$ and both $x = 0$ and $y = 0$, then $\mathit{prio}(z) = \mathit{max}(\mathit{prio}(x), \mathit{prio}(y))$.
\item If $z$ is 1, then both $x = 1$ and $y = 1$ and therefore $\mathit{prio}(z) = \mathit{min}(\mathit{prio}(x), \mathit{prio}(y))$.
\end{enumerate}
After propagating, we pick the lowest priority, say $\mathit{prio}(v_0)$, which arrived at some next state variable from $d'$.
The propagation of $\mathit{prio}(v_0)$ to a next state variable from $d'$ means that we could not
avoid to include $v_0$ into the implicitly constructed system of justification paths, although we
prefer variables with high priority. Thus we add the according literal of $v_0$ from $m$ to our
(initially empty) generalized proof obligation cube $\hat{c}$.
We now set $\mathit{prio}(v_0) = \infty$, because we already consider this variable in our generalized cube,
and start with the next forward propagation iteration.
We terminate, once we only observe priorities $\infty$ at the next state variables after propagating the current priority assignment. Then we include all corresponding literals into the partial assignment.
}

A partial assignment achieved by this method is 01X-simula\-table.

\subsubsection{Lifting}\label{sec:chocklerlifting}
The authors of~\cite{Choc:2011} propose an approach which uses an unsatisfiable SAT solver query that reveals
a generalization of the \po{} state.
Assume a circuit defining a transition function $T$.
In the original PDR approach a satisfiable query $\txt{SAT?}[\neg d \wedge R_{k-1}\wedge T\wedge d']$
provides a satisfying minterm $m$ and some complete assignment $i$ to the primary inputs $\v{i}$.
Since $m \wedge i$ is a complete assignment to all inputs of the circuit defining
the transition function, it implies a fixed next state in the cube $d'$.
Thus, the `lifting query' $\txt{SAT?}[m \wedge i \wedge T \wedge \neg d']$
is unsatisfiable by construction.
The final conflict clause of this query yields a generalization of $m$, because we are now able to remove all literals from $m$ which are unnecessary for the unsatisfiability proof.
Again results are not necessarily optimal, since they depend on the order in which the literals of $m$ propagate during the SAT solving (and since only a fixed input assignment $i$ is considered).
To further increase the number of removed literals in lifting, \cite{Ravi:2004} proposes to iteratively omit literals from the unsatisfiable core (revealed by a final conflict clause) and query the solver again with the corresponding unsatisfiable lifting call. This procedure is called \textit{literal dropping} and trades runtime against more general POs. For our experiments in Sect.~\ref{ssse:origpdr_hwmcc1517} we consider both variants.

\changes{\subsection{Limitations and Extensions of Lifting}\label{sse:lifting_ext}}
\changes{Here we discuss the preconditions we require for a sound application of lifting in PDR as well as possible extensions in order to improve its efficiency.}
For the lifting approach to be correct, $T$ has to represent a \emph{function} (\ie{} $T$ is \emph{left-total} and \emph{right-unique}, see Sect.~\ref{sse:basics}). We now consider those two properties separately.

\begin{figure}
\centering
\includegraphics[scale=0.6]{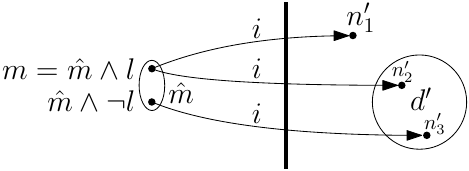}
\caption{Not right-unique}
\label{fig:Lifting_rightunique}
\end{figure}
Firstly, we assume that $T$ is \emph{not} right-unique (see Fig.~\ref{fig:Lifting_rightunique}). This means that the
assignment $m \wedge i$ does not necessarily imply \emph{one unique} successor state.
This property could render our lifting query $\txt{SAT?}[m \wedge i \wedge T \wedge \neg d']$ satisfiable,
since there could indeed be another transition from $m \wedge i$ to a state outside $d'$.
Thus, the approach would say that the \po{} $m$ cannot be generalized, although
this could actually be possible.
Existentially quantifying the input vector $\v{i}$ instead of setting it to one fixed assignment $i$
would not improve the situation (but rather make it worse), because this would increase the
probability of having transitions to states outside $d'$.

\begin{figure}
\centering
\includegraphics[scale=0.6]{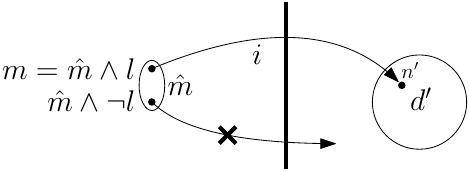}
%\vspace{-5mm}
\caption{Not left-total}
%\vspace{-2mm}
\label{fig:Lifting_lefttotal}
\end{figure}

Secondly, we assume that $T$ is \emph{not} left-total (see Fig.~\ref{fig:Lifting_lefttotal}). This means that there
are present state / input combinations which do not lead to any successor state at all. We consider the state $\hat m$ which results from removing literal $l$ from $m$, \ie{} $\hat m = m \setminus \{l\}$. Thus in the beginning $\hat m \wedge l \wedge i \wedge T \wedge d'$ is satisfiable.

We further assume that $\hat m \wedge \neg l$ has no successor in $T$ at all ($T$ is not left-total), \ie{} $\hat m \wedge \neg l \wedge T$ is already unsatisfiable. Now $\hat m \wedge i \wedge T \wedge \neg d'$ is unsatisfiable, such that
the lifting query would remain unsatisfiable when literal $l$ is dropped from $m$. However, $\hat m$ is not necessarily a correct \po{}, since it is not possible to reach $d'$ from each point (state) in $\hat m$.

To summarize, while missing right-uniqueness can only lead to (unnecessarily) failing lifting attempts, it is most crucial to ensure left-totality, since otherwise lifting could lead to wrong results \changes{in terms of spurious counterexamples}.

Finally, we discuss two variants which can potentially speed up the lifting approach
and/or improve its results.
\changes{
We apply the approximate SAT approach from \cite{Grig:2016} to lifting in order to investigate its isolated effect on \po{} generalization.
We further introduce literal rotation from \cite{Sche:2021} to bit-level SAT-based PDR.
}

\subsubsection{Approximate SAT}
If we decide to apply iterative literal dropping, we can make use of the observation made in \cite{EMB:2011} that
\changescs{-- in the context of lifting --}
satisfiable calls of the SAT solver are much more costly in terms of runtime than unsatisfiable calls and that the SAT solver usually reports unsatisfiability after only deciding few variables.
The authors of \cite{Grig:2016} therefore propose a technique which they call `approximate SAT' and which considers any SAT call as satisfiable once a certain number of decisions is made by the SAT solver (in \cite{Grig:2016} a constant number of 100 is proposed).
Hence, we can avoid unnecessary computation time in satisfiable SAT solver calls which would in the end
only conclude that we have to keep a certain literal anyway.
On the other hand we could prematurely conclude that a call is satisfiable and keep a literal even though it would not have been necessary.
Thus there is a trade-off between runtime and accuracy.
In our experimental section we will analyze whether this technique is worthwhile.

\subsubsection{Literal Rotation}
\label{sec:pogen:litrot}
The authors of \cite{Sche:2021} propose to additionally `rotate' literals in order to replace or complement standard literal dropping.
Technically, we provide the cube $m = l_1 \wedge \ldots \wedge l_k \wedge \ldots \wedge l_n$
as assumptions to an incremental SAT solver \changescs{(in the order $l_1, \ldots, l_n$)}.
Once a literal $l_k$ is conflicting, the SAT solver will traverse the implication graph and collect the previously decided assumptions $l_i, \ldots, l_j$ with $i,j \in\{1,\ldots,k-1\}$ which are necessary for the conflict (unsatisfiable core).
lifting without any literal dropping would conclude that the literals $l_i, \ldots, l_j, l_k$ are necessary and subsequently, that we may generalize $m$ to $\hat{m} = l_i \wedge \ldots \wedge l_j \wedge l_k$.
Literal \emph{rotation} however invokes the SAT solver again on the reduced set of assumptions $\hat{m}$ and `rotates' the order of the assumptions to $(l_k, l_i, \ldots , l_j)$.
As a result a newly found unsatisfiable core will at most contain all literals of $\hat{m}$ and will
\changescs{possibly}
reveal a more general unsatisfiable core.
We remark that since $l_i, \ldots , l_j$ obviously implies $\neg l_k$ the call will remain unsatisfiable.
These solver queries are rather inexpensive \cite{Sche:2021} and therefore, we can repeat rotating until some literal would re-appear as
the first one in the order.
We remark, that even if we have performed all possible rotations on an initial conflict, there might still be literals in the resulting cube which can be removed by further literal dropping \cite{Sche:2021}.
Again, we will discuss the efficiency of this method in our experimental section.

\changes{\subsection{Implication Graph Based Generalization (IGBG)}\label{sec:impl}}
We can also adapt another method to PO Generalization in PDR which is inspired from~\cite{Ravi:2004}
as the aforementioned lifting method.
Having a circuit, 
applying a full assignment $i$ (for primary inputs) and $m$ (for state variables) to $T$, \ie{} querying the SAT solver with $\txt{SAT?}[m \wedge i \wedge T]$, the SAT solver will only require
\booleanword{} Constraint Propagation (BCP) to deduce a satisfying assignment of the formula. Hence it is possible to just traverse the implication graph in a backward direction and collect the literals from $m$ which are responsible for implying the next state valuation $d'$. Obviously, the method makes use of the right-uniqueness property of transition functions (since otherwise BCP would not be sufficient).

Interestingly, a reduced cube $\hat{c}$ resulting from this method is exactly 01X-simula\-table, \ie{} if we apply it as a simulation pattern with all don't care literals (which are not contained in $\hat{c}$) set to $X$, then no $X$-value will propagate to the next-state variables included in $d'$~\cite{Ravi:2004}.

\changes{\subsection{MaxSAT 01X-encoding}\label{sse:ms01x}}
In order to avoid the iterative greedy approach of 01X-simulation for removing redundant state bits,
we introduce a partial MaxSAT~\cite{fu2006solving, li2009maxsat, ansotegui2013sat} encoding to find a better approximate solution to POGP.
Partial MaxSAT problems consist of hard clauses and soft clauses. A MaxSAT solution satisfies
all hard clauses and a maximal number of soft clauses.

For the 01X-encoding of the \booleanword{} circuit for the transition function
we introduce \emph{two} variables $v^{(0)}$ and $v^{(1)}$ for each \booleanword{} variable $v$ which represents either an input, an output or an internal signal, while $((v^{(0)} = 0 \wedge v^{(1)} = 0) \leftrightarrow v = X)$ as well as $((v^{(0)} = 1 \wedge  v^{(1)} = 0) \leftrightarrow v = 0)$ and $((v^{(0)} = 0 \wedge v^{(1)} = 1) \leftrightarrow  v = 1)$; we explicitly forbid $(v^{(0)} = 1 \wedge v^{(1)} = 1)$. All gates are replaced by a two-rail encoding according to~\cite{JBM+:2000}.
The 01X-encoded circuit simulates information propagation using 01X-logic.
For each state variable $s_i$ we introduce a new variable $t_i$ and a unit soft clause $sc_i = \{t_i\}$ accompanied by the hard clauses representing $t_i \leftrightarrow ((s_i^{(0)} = 0) \wedge (s_i^{(1)} = 0))$.
Starting with a satisfying solution to $\txt{SAT?}[\neg d \wedge R_{k-1} \wedge T \wedge d']$
with
$d' = ((d'_{i_1})^{\tau_1} \wedge \ldots \wedge (d'_{i_k})^{\tau_k}$) that
provides
full assignments $m = s_1^{\sigma_1} \wedge \ldots \wedge s_m^{\sigma_m}$ and
$i = i_1^{\iota_1} \wedge \ldots \wedge i_n^{\iota_n}$, we introduce
hard clauses fixing state bits $s_i$ to $X$ or $\sigma_i$,
input bits $i_j$ to $\iota_j$, and
next state bits $d'_{i_j}$ to $\tau_j$.
The other hard clauses of the considered MaxSAT problem correspond to the 01X-encoding
of $T$.
Maximizing the number of satisfied soft clauses
means maximizing the number of present state bits which are assigned to $X$ and are thus not included in the
resulting \subcube{} $c$ of $m$ from which all transitions under $i$ lead into $d'$.
We call the resulting MaxSAT problem \emph{MS01X}.

\changes{\subsection{SAT 01X-encoding}\label{sse:ternsat}}
 It is also possible to approximate \emph{MS01X} by using a simple SAT solver \cite{Fran:2010}. We compute a 01X-encoding of the circuit like we do for \emph{MS01X}, but omit the MaxSAT-specific
clauses. Here the notion of \emph{sign-minimality} \cite{Roor:2006} is exploited, which describes the fact that if a SAT solvers' decision heuristics only decides Boolean variables with one polarity (say 0), then the resulting model has a (locally) maximal number of variables assigned to this polarity. If we employ the 01X-encoding scheme from above
(which encodes $X$ with (00)) and
the SAT solver makes only decisions to 0, then the resulting model has a
(locally) \emph{maximal} (but not necessarily globally \emph{maximum})
number of state bits assigned to $X$.
In the following, we call this technique \emph{S01X}.

\changes{\subsection{PO Generalization with Invariant Constraints}\label{sec:selfloops}}
It is important to note that even in the context of transition relations defined by circuits,
the transition relation is not necessarily a function.
A common reason for non-left-total transition relations are invariant constraints (\eg{} restricting the inputs).
The AIGER 1.9 standard~\cite{Bier:2011} is a popular example for this.
Here a circuit with transition relation $T$ is restricted by
an invariant constraint $C$.
If some present state / input combination does not satisfy
$C$,
then there is no
transition from this state under this input assignment, \ie{} the resulting transition relation is not
left-total. This immediately implies that the lifting approach to PO generalization may produce erroneous
results (see above) and cannot be used.
There are several options to avoid this problem:
\begin{enumerate}
\item One can use PO generalization techniques for general transition relations that will be discussed
in Sect.~\ref{sse:general}. Our experimental results show however that this leads to sub-optimal generalizations.
\item One can use \emph{IGBG}
\changes{which} requires right-uniqueness to be applicable, but not left-totality.
\item One can use 01X-simulation with the additional requirement that it does not produce an $X$ at the
output of $C$.
\item It is possible to transform the transition relation into a right-unique and left-total transition function.
\begin{enumerate}
\item  We can maintain the same set of reachable states by introducing self-loops for each non-admissible transition. For this, we simply introduce for each state variable a multiplexer which feeds back the old state value in case that the invariant constraint is violated.
\item 
We can introduce a new dead-end state and direct all non-admissible transitions into this state.
To do so, the tool \emph{aigunconstraint} from the AIGER-suite~\cite{Bier:2011} introduces an additional latch
for implementing the dead-end state (doubling the state space and thus changing the set of reachable states).
\end{enumerate}
\item
The original lifting call from Sect.~\ref{sec:chocklerlifting} can be changed to consider the invariant constraints. If we are able to separate $C$ from $T$ having
$T = \hat{T} \wedge C$,
we can change $\txt{SAT?}[m \wedge i \wedge T \wedge \neg d']$ into $\txt{SAT?}[m \wedge i \wedge \hat{T} \wedge (\neg C \vee \neg d')]$. By construction, the minterm $m$ satisfies the invariant constraint and the transition from $m$ under $i$ leads into $d'$. Therefore the changed SAT query is unsatisfiable as well.
If the SAT query remains unsatisfiable for some subcube $c$ of $m$, then it is of course unsatisfiable for each state $m'$ in $c$,
thus each such $m'$ satisfies $C$ and the transition from $m'$ under $i$ leads into $d'$, \ie{} $c$ is a \po{}.
\end{enumerate}

\changes{\section{PO Generalization for General Transition Relations}\label{sse:general}}

Here, we look into methods for PO generalizations that work without specific assumptions on the
transition relation. We start with approximate techniques and finally consider exact solutions.
First, we describe a new technique called GeNTR.
We have already used a corresponding technique in the
context of SMT-based PDR \cite{Sche:2021}.
Secondly, we adapt well-known covering approaches and use them in the context of bit-level PDR. Lastly, we introduce techniques which solve the underlying 2-QBF problem and remove literals greedily as well as optimally using MaxQBF. While both QBF approaches are completely new in the context of PDR, a similar MaxQBF approach has been used in \cite{Saue:2013} in the context of test cube generation.

\changes{\subsection{Generalization with Negated Transition Relation (GeNTR)}}
If $T$ does not represent a function,
it is possible to lift $m$ with a similar formula as in the lifting approach from Sect. \ref{sec:chocklerlifting}.
Assume a satisfiable query $\txt{SAT?}[\neg d \wedge R_{k-1}\wedge T\wedge d']$.
A SAT solver provides complete assignments $m$, $t'$, and $i$  to state variables $\v{s}$,
next state variables $\v{s}'$, and additional variables
$\v{i}$ of $T$.
The cube \mbox{$m \wedge i \wedge t'$} represents a
satisfying assignment of the predicate $T$.
Hence, this cube renders $\neg T$ unsatisfiable.
Therefore, the query
\mbox{$\txt{SAT?}[m \wedge i \wedge \neg T \wedge t']$} is  unsatisfiable and its final conflict
clause can be used to obtain a generalization of $m$.

\changes{\subsection{Covering Approach}\label{sec:cover}}
Another technique for general transition relations 
is the extraction of a minimal satisfying assignment given a complete satisfying assignment from a SAT solver query. Extracting a partial assignment which still satisfies all clauses is equivalent to the \emph{Hitting Set} problem, a special case of \emph{Set Cover}~\cite{Karp:1972}.
We give the intuition: Given a set of clauses $\Gamma$ and a full satisfying assignment $A$, pick a subset of the elements (literals) of $A$ which `hit' all  clauses in $\Gamma$.
For brevity, we focus on the most commonly referenced methods in this context -- a greedy algorithm and an ILP-encoding.
\paragraph{Greedy Algorithm}
We start with a full satisfying assignment $A$ to
$\neg d \wedge R_{k-1}\wedge T\wedge d'$ and $\Gamma$ contains the clauses
representing $T$.
Initially, our partial assignment $P$ consists of all literals in $A$ which are not present state literals.
$P$ is removed from $A$, and all clauses which are covered by literals from $P$ are removed from $\Gamma$.
Then we (1) scan the clauses of $\Gamma$ for the most frequently occurring literal $l$ of $A$, (2) add $l$ to the partial assignment $P$, remove $l$ from $A$, remove the clauses covered by $l$ from $\Gamma$, and start over with (1) until $\Gamma$ is the empty set. Obviously, the greedy algorithm has a polynomial runtime in the input size.
However, the solution is not necessarily optimal \wrt{} the size of the partial assignment.

\paragraph{ILP-encoding}
The ILP encoding has a binary ILP-variable $v_l$ for each present state variable $s_i$ with literal $l$ occurring in $A$.
It
is formulated in a way that in the solution
$v_l$ equals $1$ iff the corresponding literal $l$ (from $A$) occurs in the covering.
The optimization goal is to minimize the sum over all ILP-variables $v_l$.

The two (unate) covering approaches mentioned above start with complete assignments
$\v{\sigma}$, $\v{\tau}$, and $\v{\iota}$  to state variables $\v{s}$,
next state variables $\v{s}'$, and additional variables
$\v{i}$ of $T$, keep the assignments $\v{\iota}$ and $\v{\tau}$ fixed and minimize the
remaining assignments in $\v{\sigma}$ while still satisfying $T$.
To give the cover approach a higher degree of freedom, we
can also allow to vary the assignments $\v{\iota}$ to $\v{i}$ and
$\v{\tau}$ to $\v{s}'$ as long as $\v{\tau}$ remains in the next state cube $d'$.
This additional degree of freedom could be easily integrated into the ILP formulation.
However, with this additional degree of freedom we rather consider an \emph{approximate} approach based on a SAT solver.

\paragraph{SAT-based Cover}
In the spirit of the \emph{S01X} approach for circuits from Sect.~\ref{sse:ternsat}
we introduce for each present state variable $s_i$ two new variables $s^{(0)}_i$ and
$s^{(1)}_i$. An assignment of $(1,0)$, $(0,1)$, or $(0,0)$ to $(s^{(0)}_i, s^{(1)}_i)$
means that $s_i$ is $0$, $1$ or unassigned ($X$), resp..
We replace in the CNF for $T$ all occurrences of $s_i$ / $\neg s_i$
by $s^{(1)}_i$ / $s^{(0)}_i$ as well as we add a
clause $\{\neg s^{(0)}_i, \neg s^{(1)}_i\}$ to rule out the illegal value $(1,1)$ \cite{Roor:2006,Fran:2010}.
To ensure that $s_i$ can only be unassigned or equal to the value $\sigma_i$
fixed by $m = s_{1}^{\sigma_{1}} \wedge \ldots \wedge s_{m}^{\sigma_{m}}$ we assign
$s_i^{(1-\sigma_i)}$ to 0. Moreover, we assign next state variables to enforce
$d'$. As already discussed in Sect.~\ref{sse:ternsat}, a SAT solver which only decides
variables to 0 then computes a solution with a locally maximal number of unassigned present
state variables.

\changes{\subsection{Solving the QBF Problem}\label{ssse:QBF}}
As stated in Sect. \ref{sec:complexity} the problem of PO generalization is inherently a 2-QBF problem, thus for achieving a minimal-sized \po{} we have to solve a QBF problem.
However, the QBF formulation is just a \emph{decision problem} which checks whether each state in a \emph{given} cube $c$
has a transition into a cube $d'$. Reducing a given minterm $m$ over state variables to a minimal
cube $c$ is an \emph{optimization problem} for which we consider two options:
\paragraph{Greedily applying a QBF solver}
With this approach we iteratively probe single state literals for don't cares. Instead of flipping a variable to $X$ and simulating the circuit, this generalized approach probes a state variable by universally quantifying it.
Assume that
we start with a minterm $m$ which is a full assignment to the state variables.
If we want to check whether the variables from $\v{r}$ can be removed from $m$, leading to $\hat{m}$,
we give the query $\txt{SAT?}[\forall \v{r} \exists \v{k} \exists \v{i} \exists \v{s}' : \hat{m} \wedge T \wedge d']$ to the QBF solver. Here $\v{k}$ are the variables from $\v{s}$ remaining in $\hat{m}$.

\paragraph{Applying MaxQBF}
To achieve an optimum we even have to go one step further than pure QBF solving. The notion of (partial) MaxQBF~\cite{Igna:2013, Reim:2014} allows us to add soft clauses to our problem to find the maximum number of removable literals. Similar to our MaxSAT approach from above, we introduce a soft clause per state variable which is a candidate for removal. The encoding is more complicated though, since we have to maximize the number of universally quantified state variables. We adapt a technique from~\cite{Reim:2014} which uses a multiplexer for each state variable selecting between either the assignment from $m$ or a universally quantified variable:
%%cs The the select input of the multiplexer is connected to a new variable $u_i$.
\begin{compactitem}
\item For each state variable $s_i$ we introduce
    a variable $s_i^{\forall}$ as well as a variable $u_i$.
\item We add clauses for $C_i^{\forall} = u_i \rightarrow (s_i \leftrightarrow s_i^{\forall})$ and $C_i^{\varepsilon} = \neg u_i \rightarrow (s_i \leftrightarrow \varepsilon)$ where $\varepsilon$ is the original assignment to $s_i$ in $m$.
\item Furthermore we introduce a unit soft clause $\{u_i\}$ for all $s_i$.
\item We solve the MaxQBF problem $\exists \v{u} \forall \v{s}^{\forall} \exists \v{s} \exists \v{i} \exists \v{s}' : T(\v{s},\v{i},\v{s'}) \wedge d' \wedge \bigwedge_{i=0}^n C_i^{\forall}  \wedge \bigwedge_{i=0}^n C_i^{\varepsilon} \wedge \bigwedge_{i=0}^n u_i$.
\end{compactitem}
If in the solution to
the MaxQBF problem
the soft clause $\{u_i\}$ corresponding to state variable $s_i$ is satisfied,
then $s_i$ may be universally quantified and therefore removed from the \po{}. The result provides a
minimal \subcube{} $c$ of $m$ such that there are transitions from all states in $c$ to $d'$.

\changes{\section{Analysis And Improvements}\label{sec:gencap_impr}
In Sect. \ref{sec:gencap}, we show how to categorize the different approaches we discussed in the preceding sections.
Sect.~\ref{sec:specrev}
analyzes the effects of methods from Sect.~\ref{sse:general} on the special case of left-unique transition relations.
Furthermore, we discuss the combination of certain techniques in order to achieve stronger generalization results in Sect.~\ref{sse:comb}.
Finally, in Sect.~\ref{sse:morefreedom}, we analyze the additional degree of freedom we might achieve by neglecting the restriction to an initial proof obligation consisting of a minterm (full satisfying assignment of the state variables).}

\changes{\subsection{Categorizing Generalization Capabilities \label{sec:gencap}}}

\begin{figure}
	\centering
	\includegraphics[scale=0.55]{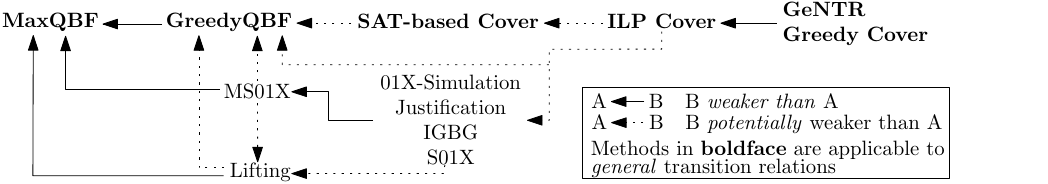}
	\caption{Quality relations between generalization methods}
	\label{fig:genDomGraph}
\end{figure}

Fig.~\ref{fig:genDomGraph} gives an overview of the methods presented in Sects.~\ref{sec:circuitmethods} and
\ref{sse:general}. The methods in the first line (written in bold) are suitable for general transition relations,
the other methods need transition \emph{functions} or at least a property implied by the
function property (such as left-totality for lifting and right-uniqueness for \emph{IGBG}).
In the following, we classify the different methods \wrt{} their generalization strength.
A solid arrow from method $B$ to method $A$ means that method $B$ is weaker than $A$ in the sense
that $B$ \changes{always} computes weaker \changes{(or equivalent)} generalizations.
\changes{
\begin{definition}[Weaker Generalization]
Consider arbitrary POGPs consisting of present state minterms (\po{} states) $m$, transition relations $T$, and next state cubes $d'$.
Assume that method $A$ generalizes $m$ to cube $c^A$, whereas method $B$ generalizes $m$ to cube $c^B$.
We consider $B$ \emph{weaker} than $A$, if $c^A$ contains always \emph{less or equally many} literals than $c^B$, i.e., $c^A$ is more or equally general than $c^B$.
\end{definition}

In many cases the relation between two methods $A$ and $B$ is not that easy to categorize.
Methods for \po{} generalization often use heuristics (for removing literals in certain orders, e.g.)
and depending on heuristical choices the results are better or worse
- so the methods are often incomparable.
Nevertheless we introduce here the notion of `potentially weaker'
(represented by dashed arrows in Fig.~\ref{fig:genDomGraph} from method $B$ to method $A$ when
$B$ is `potentially weaker' than $A$). This notion intuitively means that method $B$ has a
conceptual weakness compared to method $A$
(like 01X-simulation which suffers from the imprecision of 01X-logic, in contrast to lifting). This weakness may
or may not be hidden by different heuristical decisions.
For a formal definition of `potentially weaker' we make use of the fact that
the methods for \po{} generalization do not need to be started with minterms, but they can
also be started with cubes which then either cannot be confirmed to be \pos{} or can be confirmed (and possibly even improved).
When $B$ is `potentially weaker' than $A$, then $B$ \emph{may} produce more general (better) results than $A$
(due to different heuristical decisions), but
if we apply $A$ to a result of $B$, then $A$ will always at least confirm the result or even improve it.
(E.g. lifting started with result cubes of 01X-simulation can always guarantee that those result cubes are valid \pos{},
but 01X-simulation cannot always give such a guarantee for cubes computed by lifting due to the imprecision of 01X-logic.)

\begin{definition}[Confirmation of a Generalization Result]
Consider an arbitrary POGP consisting of a \po{} state $m$, a transition relation $T$, and a next state cube $d'$.
Assume that
$m$ is generalized to a cube $c^B$ by method $B$.
Then method $A$ is able to \emph{confirm the generalization result} of method $B$,
if started with $c^B$, $T$, and $d'$, it is able to decide that $c^B$ is a valid \po{} or if it is able to generalize $c^B$ even further.
\end{definition}

\begin{definition}[Potentially Weaker Generalization]
If method $A$ is able to confirm the generalization result of method $B$ for all POGPs, then $B$ is \emph{potentially weaker} than $A$.
\end{definition}
}

We first look into the general methods.
Beforehand, we prove that the \emph{GeNTR} approach implicitly computes a cover of the CNF for $T$.
\begin{restatable}{theorem}{gentrtheo}
\label{thrm:gentrcover}
\emph{GeNTR} (implicitly) computes a cover of the CNF for $T$. 
\end{restatable}
\begin{IEEEproof}
\emph{GeNTR} starts with complete assignments $m$, $t'$, and $i$  to state variables $\v{s}$,
next state variables $\v{s}'$, and additional variables $\v{i}$ of $T$. The query
\mbox{$\txt{SAT?}[m \wedge i \wedge \neg T \wedge t']$} is unsatisfiable.
It computes a \subcube{} $c$ of $m$ such that
\mbox{$\txt{SAT?}[c \wedge i \wedge \neg T \wedge t']$} is still unsatisfiable, \ie{}
each extension of $c \wedge i \wedge t'$ to a full assignment evaluates $\neg T$ to 0 and thus
$T$ to 1.
This means, that each such extension satisfies all clauses in $T$.
This implies that $c \wedge i \wedge t'$ has to cover all clauses in $T$.
%\qed{}
\end{IEEEproof}

Thus, \emph{GeNTR} is equivalent to the greedy cover approach.
Of course, the results may differ, since different heuristics are used.
The ILP cover approach %%cs ILP$^{\text{fix}}$
solves the covering problem in an optimal way and is thus
stronger than greedy covering and \emph{GeNTR}.
Nevertheless, it is `potentially weaker' than \emph{SAT-based cover}, since
\emph{SAT-based cover} has the freedom to choose different values for $\v{i}$ variables
and $\v{s}'$ variables which are not fixed by $d'$.
As already discussed in Sect.~\ref{ssse:QBF} QBF is exact as a decision problem, but
it is not an optimization method which is able to \emph{compute} the reduced cube.
Greedy application of QBF depends
on the chosen order and therefore it is only potentially stronger than
\changes{all other methods except for MaxQBF}.
MaxQBF provides an optimal solution and dominates all other methods.

Now we look at methods suitable for transition functions.
\emph{IGBG}, \changescs{\emph{justification}}, 01X-si\-mu\-la\-tion, and \emph{S01X} have basically the same strength, since the results of \emph{IGBG} \changescs{and \emph{justification} are}
01X-simulatable~\cite{Ravi:2004} and the result of \emph{S01X} is only locally optimal.
The
\changescs{four}
methods may produce different $X$-values on present state variables, however.
01X-simulation, \emph{IGBG}, \changescs{\emph{justification}}, and \emph{S01X} are weaker than \emph{MS01X},
since MaxSAT computes an \emph{optimal} selection
of $X$-values for state bits.
All 01X-based methods are `potentially weaker' than lifting.
If $X$-values on present state variables do not propagate to the next state bits included in the
next state cube $d'$ for a fixed assignment to the remaining present state variables and the primary inputs,
then the values assigned to the $d'$-variables are implied by this assignment.
In other words, this assignment, the transition relation $T$ and $\neg d'$
are contradictory, and thus the present state bits with $X$-values can be potentially removed by the lifting method.
On the other hand lifting does not suffer from the known imprecision of ternary logic as discussed in Sect.~\ref{01X}.

\begin{figure}
	\includegraphics[scale=1.2]{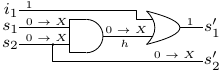}
	\caption{Example~\ref{ex:cover}.}
	\label{fig:ex2:weaknessCoverApproaches}
\end{figure}

The weaker,  covering based methods for general transition relations
(greedy cover, \emph{GeNTR}, ILP cover)
should not be preferred over circuit-based methods
(if they are applicable), since they have a fundamental weakness as illustrated by the following
example:

\begin{example}
\label{ex:cover}
Consider a transition relation $T$ specified by
the circuit in Fig.~\ref{fig:ex2:weaknessCoverApproaches}.
The next state cube $d'$ is given by $s_1'$.
We assume that a SAT solver produces the satisfying assignment
$i_1 = s_1' = 1$, $s_1 = s_2 = s_2' = 0$, \ie{} we start generalization with the minterms $m = \neg s_1 \wedge \neg s_2$,
$i = i_1$.
It is easy to see that 01X-simulation can assign both $s_1$ and $s_2$ to $X$, leading to an $X$ at output $h$
of the AND-gate, but keeping the $1$ at the output $s_1'$ of the OR-gate. The lifting approach can remove the assignments
to both $s_1$ and $s_2$ as well. However,
covering the clauses $(\neg h \vee s_1)$, $(\neg h \vee s_2)$, and $(h \vee \neg s_1 \vee \neg s_2)$
of the AND-gate can only remove \emph{one} of the input assignments (\changescs{$\neg s_1$ \emph{or} $\neg s_2$}).
In general, clause covering means to find assignments to the inputs of all gates that justify the assignments at their outputs.
This property largely restricts the potential to remove input assignments to gates by clause covering approaches.
\end{example}

Of course, apart from theoretical comparisons, an experimental evaluation (see Sect.~\ref{sec:experiments}) is crucial,
since the overall effect on the PDR algorithm is also influenced
by the effectiveness of heuristics as well as by the runtimes needed to compute generalizations.

\subsection{The Special Case of Left-unique Transition Relations}\label{sec:specrev}
All methods described in Sect.~\ref{sse:general} are sound for arbitrary transition relations
and thus also for left-unique transition relations.
Our investigations for the special case of left-unique transition relations are motivated by Reverse PDR
where the predicate for $T$ is interpreted `the other way around'. If the original transition relation results from
a circuit, \ie{} if it is (left-total and) right-unique, then the reverted transition relation
is left-unique. In fact, we can prove that
the cover methods from Sect.~\ref{sse:general} (including \emph{GeNTR})
do not result in any generalization of \pos{} for
left-unique transition relations. Only the QBF- and MaxQBF-based methods are able to generalize \pos{}
in this case.
Our covering approaches minimize the number of assigned present state variables
and still cover all clauses from $T$. Therefore we can assume \wlogword{} that
all variables from $\v{i}$ and $\v{s}'$ are assigned. Since $T$ is left-unique,
a full assignment to the variables from $\v{s}$ is then implied, \ie{} removing any
assignment of a variable from $\v{s}$ would render $T$ to be unsatisfied and we would lose
the covering property of the assignment.

\begin{example}
\label{ex:and}
Assume a very simple left-unique transition relation $s_1 = i_1 \wedge s'_1$ with CNF
$T = (\neg s_1 \vee i_1) \wedge (\neg s_1 \vee s'_1) \wedge (s_1 \vee \neg i_1 \vee \neg s'_1)$.
It is easy to see that in each covering of $T$ $s_1$ is assigned.
Assume that the next state cube is $s'_1$. For each valuation of $s_1$ there is a valuation of
$i_1$ such that $T \wedge s'_1$ is satisfied.
Hence, only the QBF approach is able to remove $s_1$ from an initial cube $s_1$ or $\neg s_1$.
\end{example}

Since QBF- and MaxQBF-based methods, \changes{which are }the only promising approaches for left-unique transition relations
considered so far, are rather expensive,~\cite{SeSc:2018} introduced a \emph{structural} approach to PO generalization
for Reverse PDR on circuits.
This structural approach basically removes state variables corresponding to the outputs
of non-constant circuit outputs with disjoint support sets.
Since it is only applicable for Reverse PDR on circuits, we omit a more detailed exposition.

\subsection{Combining State Lifting Methods}\label{sse:comb}
Some methods are a good fit for collaboration, \eg{} \emph{S01X}, 01X-simulation, \changescs{\emph{justification},} the implication graph based method \emph{IGBG}, and the MaxSAT 01X-encoding \emph{MS01X}.
\emph{S01X}, 01X-simulation, \changescs{\emph{justification},} and \emph{IGBG} (see Sect. \ref{sec:impl}) yield 01X-simulatable generalization results. Thus the result of one of these methods can be a starting point to \emph{MS01X}
which may be able to improve the generalization even further.
\emph{MS01X}
introduces a soft clause per state variable which is a candidate for removal. If we already have found a set of don't cares while using another method, we only have to introduce soft clauses for the remaining state variables. For all other variables we can introduce the don't care value as a hard clause or assumption, heavily decreasing the search space of the MaxSAT solver.
In our experiments we use a heuristics for dynamically combining \emph{MS01X} with \emph{IGBG}.
If \emph{MS01X} is successful and not much slower than \emph{IGBG}, we only use \emph{MS01X} in the future.
If \emph{MS01X} is much slower and not significantly more successful than \emph{IGBG}, we use \emph{IGBG} only.
In all other cases we use the mentioned combination where \emph{MS01X} improves on a precomputed result of \emph{IGBG}.
A similar method could be used with the MaxQBF approach, only with the difference, that MaxQBF is able to improve or at least meet \emph{every} previous result of our different methods.

\subsection{More Degrees of Freedom?}\label{sse:morefreedom}
The PO generalization methods discussed so far look for a minimal subcube $c$ of a minterm $m$
(resulting from a satisfied SAT solver call $\txt{SAT?}[\neg d \wedge R_{k-1}\wedge T\wedge d']$)
with the property that $\tilde{m} \wedge T\wedge d'$ is still satisfiable for all minterms
$\tilde{m}$ covered by $c$.
Thus those methods are restricted by the initial choice of $m$ and
starting from another minterm $\hat{m}$ could lead to a much better solution.
Some of the optimization based methods like  \emph{S01X},
\emph{MS01X}, SAT-based cover, ILP-based cover,
greedy QBF, and \emph{MaxQBF} offer themselves to let the choice of the present state variables
open in the optimization.
\changescs{One small additional detail should be considered in this context:}
In order to minimize the cube $c$ mentioned above as much as possible, the methods based on a fixed initial
minterm $m$ allow $c$
to contain states which have already been excluded from $R_{k-1}$ before, but
$c$ contains at least one new state not yet excluded from $R_{k-1}$ (at least $m$).
(This is the approach used in the literature on PDR, see \eg{} \cite{Brad:2011,EMB:2011}).
However, if we let the choice of the present state variables open and only require
that $\tilde{m} \wedge T\wedge d'$ is satisfiable for all minterms
$\tilde{m}$ covered by $c$, then it could happen that $c$ contains
\emph{only} states which have already been excluded from $R_{k-1}$, so it could be useless.
Therefore we now require that $\tilde{m} \wedge R_{k-1}\wedge T\wedge d'$ is satisfiable for all minterms
$\tilde{m}$ covered by $c$, \ie{} we replace $T$ by $R_{k-1}\wedge T$ in those methods.

We call the resulting methods \emph{S01X}$^{\text{free}}$, \emph{MS01X}$^{\text{free}}$ etc.
to differentiate them from the methods considered so far which are now called
\emph{S01X}$^{\text{fix}}$, \emph{MS01X}$^{\text{fix}}$ etc.
Here we give more details on how to realize the approaches mentioned in Sect. VI-D
where the choice of the present state variables is open.

%\vspace{-2mm}
\subsubsection{\emph{MS01X}$^{\text{free}}$ and \emph{S01X}$^{\text{free}}$}
Compared to \emph{MS01X}$^{\text{fix}}$ (see Sect. IV-D) the MaxSAT encoding
has to be modified in a straightforward manner:
The soft clauses $sc_i = \{t_i\}$ and the corresponding hard clauses representing
$t_i \leftrightarrow ((s_i^{(0)} = 0) \wedge (s_i^{(1)} = 0))$ remain unchanged.
We omit the hard clauses fixing present state bits
$s_j$ to $X$ or $\sigma_j$.
The hard clauses for primary inputs are now
for all $1 \leq j \leq n$ $((i_j^{(0)} \vee i_j^{(1)})$
(only disallowing $X$ for $i_j$).
For all $1 \leq j \leq k$ the hard clauses for $((d'^{(0)}_{i_j} = (1-\tau_j)) \wedge (d'^{(1)}_{i_j} = \tau_j))$
(fixing $d'_{i_j}$ to $\tau_j$) remain unchanged.
The hard clauses representing an 01X-encoding of $T$ remain unchanged as well.
Finally, we replace in the CNF $R_{k-1}$ each literal $s_i$ by $s_i^{(1)}$ and each literal
$\neg s_i$ by $s_i^{(0)}$. We add the resulting clauses to the MaxSAT problem.
In that way we enforce that the chosen assignments to the present states are included in
of $R_{k-1}$.

As for  \emph{S01X}$^{\text{fix}}$, \emph{S01X}$^{\text{free}}$ simply results from \emph{MS01X}$^{\text{free}}$
by omitting the soft clauses $sc_i = \{t_i\}$ (and their corresponding hard clauses $t_i \leftrightarrow ((s_i^{(0)} = 0) \wedge (s_i^{(1)} = 0))$).

%\vspace{-2mm}
\subsubsection{\emph{SATCover}$^{\text{free}}$}
Compared to \emph{SATCover}$^{\text{fix}}$ (see Sect. V-Bc)
we remove the restriction that present state bits
$s_i$ can only be unassigned or equal to the value $\sigma_i$
fixed by $m = s_{1}^{\sigma_{1}} \wedge \ldots \wedge s_{m}^{\sigma_{m}}$, \ie{}
we just have to remove the assignment of $s_i^{(1-\sigma_i)}$ to 0.

%\vspace{-2mm}
\subsubsection{\emph{ILP}$^{\text{free}}$}
Compared to \emph{ILP}$^{\text{fix}}$ (see Sect. V-Bb) it is also possible to
arrive at \emph{ILP}$^{\text{free}}$ if we do not confine ourselves to the assignment $A$ but much rather
allow the ILP solver to choose between both polarities of variables.

For ILP$^{\text{free}}$ we introduce two binary ILP-variables $v_l$ and $v_{\bar{l}}$ for each variable $v$ in
$\neg d \wedge R_{k-1}\wedge T\wedge d'$, one ILP variable for each literal.
$\Gamma$ contains the clauses representing $\neg d \wedge R_{k-1}\wedge T\wedge d'$.
We introduce a linear constraint $\sum_{i=1}^n v_{l_i} \geq 1$ for each clause $c = \{l_1, \ldots, l_n\}$ of
$\Gamma$. Now we have to assert that only one polarity for each variable is used by introducing
the linear constraints $v_{l} + v_{\bar{l}} \leq 1$.
As an optimization goal we minimize the sum over all ILP-variables corresponding to present state variables.
If in the solution to ILP$^{\text{free}}$ both variables corresponding to literals of a present state variable $s_i$
are assigned to 0, then the computed generalized cube $c$ does not contain a literal for $s_i$.
The \emph{binate covering} ILP-problem now has to solve the original SAT-problem again, but with an
additional optimization component, and is therefore more complex. Nevertheless it may find a smaller partial assignment.

%\vspace{-2mm}
\subsubsection{\emph{GreedyQBF}$^{\text{free}}$}
As \emph{GreedyQBF}$^{\text{fix}}$ (see Sect. V-Ca), \emph{GreedyQBF}$^{\text{free}}$
partitions the present state variable into a set of variables $\v{r}$ which are removed from the
cube representing the \po{} and a set of variables $\v{k}$ remaining in the \po{}.
Whereas in \emph{GreedyQBF}$^{\text{fix}}$ the QBF formula is
$\forall \v{r} \exists \v{k} \exists \v{i} \exists \v{s}' : \hat{m} \wedge T \wedge d'$,
we remove in \emph{GreedyQBF}$^{\text{free}}$ the fixing of the $\v{k}$ variables to $\hat{m}$.
Moreover, we use the formula $\neg d \wedge R_{k-1} \wedge T \wedge d'$ as in the original SAT problem
leading to
$\exists \v{k} \forall \v{r} \exists \v{i} \exists \v{s}' : \neg d \wedge R_{k-1} \wedge T \wedge d'$.

%\vspace{-2mm}
\subsubsection{\emph{MaxQBF}$^{\text{free}}$}
Compared to \emph{MaxQBF}$^{\text{fix}}$ (see Sect. V-Cb), in
\emph{MaxQBF}$^{\text{free}}$
we introduce a new \emph{existentially} quantified variable $s_i^{\exists}$ for one input of the multiplexer,
instead of using the original assignment $\varepsilon$ from $m$. In such a modified MaxQBF problem \emph{MAXQBF$^{\text{free}}$}
we conjoin with $\bigwedge_{i=0}^n C_i^{\exists}$ instead of $\bigwedge_{i=0}^n C_i^{\varepsilon}$  whereas $C_i^{\exists} = \neg u_i \rightarrow (s_i \leftrightarrow s_i^{\exists})$.

For \emph{MAXQBF$^{\text{free}}$} we use the formula $\neg d \wedge R_{k-1} \wedge T \wedge d'$ as in the original SAT problem, \ie{} the MaxQBF query is $\exists \v{u} \exists \v{s}^\exists \forall \v{s}^{\forall} \exists \v{s} \exists \v{i} \exists \v{s}' : \neg d \wedge R_{k-1} \wedge T(\v{s},\v{i},\v{s'}) \wedge d' \wedge \bigwedge_{i=0}^n C_i^{\forall} \wedge \bigwedge_{i=0}^n C_i^{\exists} \wedge \bigwedge_{i=0}^n u_i$.
This query yields a \po{} with a minimal number of state variables (which still satisfies $\neg d \wedge R_{k-1}$ and) which is not constrained to $m$ as a `starting point'.

\section{Experimental Results}\label{sec:experiments}

We divide the experimental results into \emph{two sections}.
In the first one\changes{, Sect.~\ref{sse:hw_model_checking},} we discuss the results on Hardware Model Checking Competition (HWMCC) benchmarks,
in the second one\changes{, Sect.~\ref{sse:ai_planning},} the results on AI Planning benchmarks of the International Planning Competition (IPC).
For each benchmark, we limited the execution time to 3,600~s and set a memory limit of 7~GB.
We used one core of an Intel Xeon CPU E5-2650v2 with 2.6\,GHz. We provide our binaries and results under \cite{SWSS:2021}.\footnote{
\changescs{If the paper will be accepted, we will also share the sources of our implementation
with the scientific community to provide a broad basis for experiments with different PO generalization techniques.}
}

\subsection{Hardware Model Checking}\label{sse:hw_model_checking}
Our PDR implementation is based on \emph{ic3ref}~\cite{IC3r:2013} and augmented to support Reverse PDR too.
Unless the order by which literals are removed from \pos{} is given by the definition of the algorithm (as in greedy covering for instance), we always consider the variable order as implemented in ic3ref, which uses an activity-based order (literals which could be removed more often are preferred).
Furthermore, we leave the \emph{clause} generalization part of ic3ref which uses \emph{ctgDown} from \cite{Hass:2013} completely untouched.
All experiments have been performed on the complete benchmark set of HWMCC'15~\cite{Hwmc:2015} and '17~\cite{Hwmc:2017} excluding the access restricted Intel benchmarks (730 instances) and on
the subset of HWMCC'19~\cite{Hwmc:2019} bit-vector benchmarks containing invariant constraints (231 instances).
If a generalization technique requires a SAT solver, we stick to \textsc{MiniSat} v2.2.0~\cite{Een:2003} as used in ic3ref. As MaxSAT solver we use \emph{Pacose}~\cite{Paxi:2019}
with Glucose 4~\cite{audemard2018glucose} and DGPW encoding~\cite{Paxi:2018},
as QBF solver \emph{DepQBF}~\cite{Lons:2010} with incremental solving. The used MaxQBF solver is \emph{quantom}, an extension of \emph{antom}~\cite{Schu:2010} which is a rather experimental implementation on top of a search-based QBF solver. Finally, \emph{gurobi9.02}~\cite{Guro:2020} is used for ILP-solving.
\changes{
\paragraph*{Structure of the Section}
We structure our experiments on Hardware Model Checking benchmarks as follows.
In a first set of experiments we considered the original (forward) PDR without invariant constraints
applied to HWMCC benchmarks (see Sect.~\ref{ssse:origpdr_hwmcc1517}).
\begin{itemize}
\item In the first experiment, we compared the \emph{generalization capabilities} of all
techniques.
To enable a fair evaluation of all methods
by comparing them on exactly the same problems, we extracted \emph{single} PO generalization
problems (POGPs) from PDR runs on HWMCC benchmarks (see Sect.~\ref{para:pogp_single}).
\item Next, we analyze the different methods within full PDR runs (see Sect.~\ref{para:full_pdr} and Sect.~\ref{para:full_pdr_ratios}).
\end{itemize}

In Sect~\ref{ssse:revpdr}, we consider the special case of left-unique transition relations (see Sect.~\ref{sec:specrev}) with the example of Reverse PDR.
The experiments refer to full PDR runs.

Finally, in Sect.~\ref{ssse:invariant_constraints}, we examine full forward (standard) PDR runs on AIGER 1.9 \cite{Bier:2011} Benchmarks with invariant constraints that invalidate left-totality.
}

\begin{table*}%[t]
    \scriptsize
	\centering

	\begin{tabular}{|c|c|c|c|c|c|c|c|c|c|c|c|c|c|c|}
		\multicolumn{2}{c|}{} & GeNTR & \shortstack{Greedy \\ Cover} & \shortstack{ILP \\Cover} & \shortstack{SAT \\Cover} & \shortstack{01X \\sim.} & S01X & MS01X & IGBG & \shortstack{Justifi-\\cation} & Lifting & \shortstack{Lifting \\+ lit.drop.} & \shortstack{Greedy\\ QBF} & MaxQBF \\
		\cline{2-15}
		\multicolumn{15}{c}{}\\[-7pt]
		\hline
		\multirow{2}{*}{\rotatebox[origin=c]{90}{fix}}%\scriptsize{fix}}}
			& Reduction & 1.07 & 1.91 & 4.36 & 4.48 & 10.76 & 10.54 & 10.84 & 9.43 & 10.76 & 9.41 & 11.58 & 14.73 & 14.85 \\
		\cline{2-15}
		& \changes{Quality} & 7.7 & 9.6 & 39.3 & 37.2 & 59.0 & 55.3 & 59.2 & 55.7 & 59.0 & 55.5 & 65.8 & 99.8 & 100 \\
		\hline
		\hline
		\multirow{2}{*}{\rotatebox[origin=c]{90}{free}}%\scriptsize{free}}}
			& Reduction & - & - & 6.09 & 4.03 & - & 8.68 & 9.96 & - & - & - & - & 12.35 & 12.68 \\
		\cline{2-15}
		& \changes{Quality} & - & - & 54.0 & 42.9 & - & 56.3 & 67.6 & - & - & - & - & 97.3 & 100 \\
		\hline
		\multicolumn{15}{c}{}\\[-7pt]
	\end{tabular}
    \caption{Reduction ratio (in \%) and \changes{quality} (in \%) for \emph{fix} and \emph{free} variants.}
	\label{tab:ratiopogp}
\end{table*}

%\vspace{-2mm}
\subsubsection{Original PDR}\label{ssse:origpdr_hwmcc1517}
\paragraph{Single PO generalization problems}\label{para:pogp_single}
For comparison, we considered only POGPs which
(a) allow for \po{} generalization and
(b) can be solved by all methods (including MaxQBF).
The problems were extracted during a MaxQBF run, \ie{}
the cubes of next state \pos{} $d'$ in
$\txt{SAT?}[\neg d \wedge R_{k-1} \wedge T \wedge d']$
are minimal. We randomly picked 258 such POGPs.
In Table \ref{tab:ratiopogp}
we present both the average reduction ratio and the average \changes{quality} of the methods.
The reduction ratio for a POGP is the number of removed state bits divided by the total number of state bits.
The \changes{quality} of a method for a single POGP relates its reduction capability to that of MaxQBF which
produces an optimal result, \ie{} the \changes{quality} of a method on a POGP
is just the quotient of the number of removed state bits for this method and the number
of removed state bits for the optimal MaxQBF
(which achieves the optimal \changes{quality} of 100\% by definition).
Table~\ref{tab:ratiopogp} shows average numbers for all 258 random POGPs.
In Table~\ref{tab:ratiopogp} we differentiate between the variants starting with
a fixed minterm $m$ from a satisfying assignment (called \emph{fix} variants)
and the variants which are not restricted by the initial choice of $m$
(called \emph{free} variants, see Sect.~\ref{sse:morefreedom}).
Interestingly, the \emph{free} variants mostly achieve worse results than their corresponding \emph{fix} variant.
The effect of the restriction that the resulting PO has to be completely included
in $R_{k-1}$ in the \emph{free} variants (see Sect.~\ref{sse:morefreedom}) apparently
outweighs the independence from the initial choice of $m$.
For the \emph{fix} variants we observe e.g. that the covering approaches are
not very suitable for HWMCC benchmarks (especially greedy cover and GeNTR).
01X-simulation has a slightly better \changes{quality}
than lifting (in spite of its theoretical inferiority due to the imprecision of ternary logic).
This is however compensated, if lifting is extended by literal dropping.
The \changes{quality} of 01X-based methods, IGBG, and lifting based methods lies between
55 and 66\% of the optimal \changes{quality} of MaxQBF which shows the potential of
more exact methods for PO generalization. Interestingly, greedy QBF achieves
already 99.8\% of the MaxQBF \changes{quality}.

\begin{figure}[t]
    \centering
  \includegraphics[width=9cm]{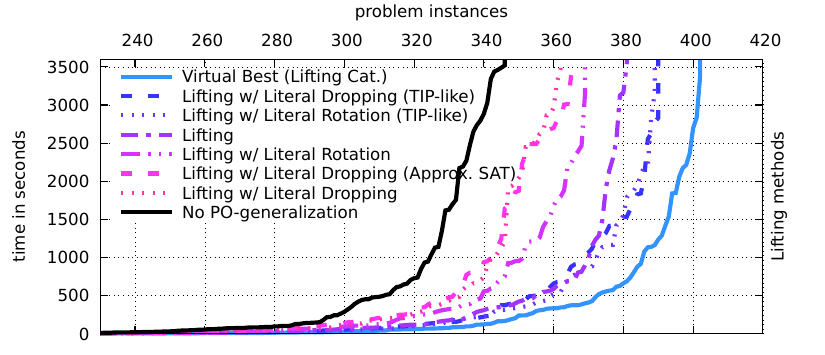}
  \includegraphics[width=9cm]{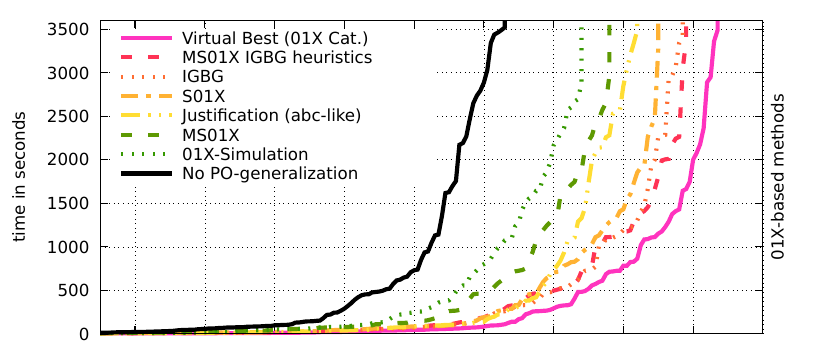}
  \includegraphics[width=9cm]{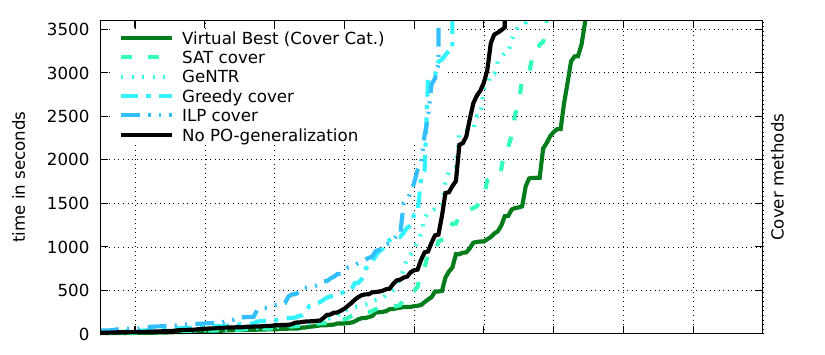}
  \includegraphics[width=9cm]{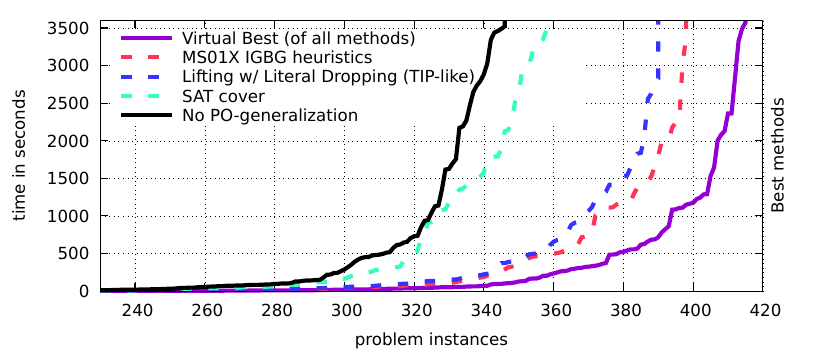}
  \caption
  {Original PDR on all techniques from 1) the lifting, 2) the 01X-based, and 3) the cover approaches. The last plot displays the best of all three categories. HWMCC'15/17 Benchmarks.
  }
  \label{fig:gen_categories}
\end{figure}

\paragraph{Full PDR runs, overall performance}\label{para:full_pdr}
Note that within full PDR runs the single POGPs become different due
to different generalization results and runtimes of the generalization
methods become relevant as well.
In Fig.~\ref{fig:gen_categories} we restrict the presentation to the methods
with the most promising performance on the set of HWMCC'15/17 Benchmarks
(in particular, we do not show any \emph{free} variants).

We start with comparing methods from each category to determine their respective strongest methods.
As categories, we consider \emph{lifting} with its different configurations,
\emph{01X-based methods} (IGBG, 01X-simulation, S01X, MS01X, and justification), and \emph{cover approaches} (ILP, SAT cover, greedy cover, GeNTR).

In the lifting category, we compare standard lifting from Sec. \ref{sec:chocklerlifting} against variants using additional literal dropping as well as literal rotation.
Since exhaustive literal dropping is very costly, we additionally try two heuristics to trade reduction ratio against runtime.
Firstly, we use approximate SAT with a decision limit of 100 -- as proposed in \cite{Grig:2016}.
Secondly, we limit the number of overall attempts to drop a literal to 32 and the number of failed attempts to 2.
After each successful attempt, we reset the current count of failed attempts to 0 and shuffle the remaining literals randomly.
We copied this strategy from TIP\footnote{Downloaded from \url{https://github.com/niklasso/tip}, Sept. 2021} \cite{EenS:2003} and therefore denote it by `TIP-like'.
Regarding literal rotation, we also implemented a version which acts `TIP-like'.
We consider one attempt of literal rotation as failed,
\changescs{if it has not been able to remove any literal from the UNSAT core.}
The first plot of Fig. \ref{fig:gen_categories} (starting at the top) displays all results from the lifting category.
Apparently, literal dropping as well as literal rotation does not pay off when performed exhaustively.
Literal rotation is performing better than literal dropping, most likely because of the relatively cheap unsatisfiable SAT solver queries (see Sect.~\ref{sec:pogen:litrot}).
Forcing \textsc{MiniSat} to report `satisfiable' after 100 decisions (approximate SAT), is able to increase the performance of literal dropping but is still not able to outperform literal rotation.
Imposing hard bounds on the number of literal dropping attempts (and fails) however, yields significantly better results.
We make the same observation for literal rotation.
The two `TIP-like' configurations of literal rotation and literal dropping are able to improve the overall performance of the `standard technique' of lifting.

The second plot of Fig. \ref{fig:gen_categories} displays all results of the 01X-based methods.
Interestingly, \emph{S01X}, the \emph{justification} approach from ABC, as well as \emph{IGBG} and the heuristics based on \emph{MS01X} / \emph{IGBG} significantly outperform the `standard technique' of greedy 01X-simulation.
\emph{IGBG} and its combination with \emph{MS01X} performs best.
Regarding \emph{S01X}, we also tried to alter the decision heuristics of \textsc{MiniSat} to decide state variables first.
However, we did not observe any relevant difference with respect to our results.

The third plot of Fig. \ref{fig:gen_categories} displays all results of the cover approaches.
\changescs{For \emph{ILP cover} and \emph{greedy cover} the relation between quality and computational cost of
PO generalization is apparently not beneficial enough}
as they perform worse than doing no PO generalization at all -- at least on Hardware Model Checking problems.
\changes{A possible reason could be its cost inefficiency which is discussed in more detail in Sect. \ref{para:full_pdr_ratios}.}
\changescs{In the category of cover approaches, PO generalization with \emph{SAT cover} solves most instances, followed by \emph{GeNTR}.}

For the results displayed by the plot at the bottom of Fig. \ref{fig:gen_categories} we picked the best performing technique from each category.
The heuristics based on \emph{MS01X} / \emph{IGBG} outperforms the best approach based on \emph{lifting}.

\changescs{In particular, it is also interesting to observe (by comparing the different plots) that
MS01X / IGBG as well as IGBG outperform the `standard techniques', i.e., lifting and greedy
01X-simulation.}
Additionally, our results indicate that we should always prefer 01X-based or lifting-based methods to cover approaches -- whenever the transition relation's properties allow for it.

\changes{The different techniques show a great variety of uniquely solved benchmarks as displayed by the overall `virtual best' at the bottom of Fig.~\ref{fig:gen_categories} as well as the `virtual best' of each category.
(The `virtual best' approach corresponds to running all methods in parallel and counting a benchmark as solved as soon as at least one method solved it.)}

\begin{table}[]
\changes{
	\centering
\scriptsize
	\begin{tabular}{c|c|c|c|c|}
				& \shortstack{No \\PO-Gen.} & \shortstack{SAT \\ Cover} & \shortstack{Lifting \\ lit.drop. (TIP)} & \shortstack{MS01X / \\IGBG heur.} \\
				\hline
				\hline
				SAT (Counterexample) & 100 & 101 & 113 & \textbf{118}  \\ \hline
                UNSAT (Safe)& 246 & 257 & 276 & \textbf{280}  \\
				\hline
	%\hline
	\end{tabular}
    \caption{\changes{Detailed results of best performing techniques from Fig. \ref{fig:gen_categories}}.}
	\label{tab:satunsat_orig}
	}
\end{table}
\changes{
We further refer to Table \ref{tab:satunsat_orig} were we split the results of the best performing techniques of each category into counterexamples (SAT) and proofs of safety (UNSAT).
Apparently, PO generalization speeds up both SAT and UNSAT.
The results also indicate, that the MS01X / IGBG heuristics is stronger than the other variants on both benchmark categories.
}

%script "calc_exectime_frac.py" - fair and reproducible ...
%script "extract_red_ratio.py" - fair and reproducible ...
\begin{table*}
	\centering
\scriptsize
	\begin{tabular}{|c|c|m{6mm}|m{6mm}|m{6mm}|m{6mm}|m{6mm}|m{6mm}|m{6mm}|m{6mm}|m{6mm}|m{6mm}|m{6mm}|m{6mm}|m{6mm}|m{6mm}|m{6mm}|}
				 \multicolumn{2}{c|}{} &\shortstack{Lifting} & \shortstack{Lifting \\ lit.drop.\\ (TIP)} & \shortstack{Lifting \\ lit.drop.} & \shortstack{Lifting \\lit.rot.\\ (TIP)} & \shortstack{Lifting\\ lit.rot.} & IGBG & MS01X & \shortstack{MS01X \\/IGBG \\ heur.} & S01X & \shortstack{Justifi-\\cation} & \shortstack{01X \\Sim.} & \shortstack{Greedy\\ Cover} & GeNTR & ILP & \shortstack{SAT\\ Cover}\\
				\hline
				\hline
				\multirow{3}{*}{\rotatebox[origin=c]{90}{\parbox{8mm}{$\%$ Exe. \\Time}}}
				 & Avg. & 2.2 & 9.4 & 45.1 & 4.0 & 24.7 & 2.0 & 60.2 & 20.5 & 5.1 & 12.4 & 46.3 & 39.6 & 8.8 & 79.7 &  6.3 \\
				\cline{2-17}%\hline
				&\changes{Median} & \changes{1.2} & \changes{3.3} & \changes{31.7} & \changes{0.8} & \changes{7.7} & \changes{1.2} & \changes{44.7} & \changes{20.5} & \changes{3.4} & \changes{3.0} & \changes{5.1} & \changes{5.8} & \changes{2.6} & \changes{84.0} & \changes{3.5} \\
				\cline{2-17}%\hline
				&\changes{Std. Dev.} & \changes{1.9} & \changes{8.6} & \changes{29.5} & \changes{6.5} & \changes{27.8} & \changes{19.2} & \changes{26.0} & \changes{22.7} & \changes{8.5} & \changes{14.5} & \changes{26.0} & \changes{11.8} & \changes{9.2} & \changes{23.9} & \changes{6.5} \\
				\hline
				\multicolumn{17}{c}{}\\[-7pt]
				\hline
				\multirow{3}{*}{\rotatebox[origin=c]{90}{\parbox{8mm}{$\%$ Red. \\Ratio}}}
				& Avg. &  50.0 & 53.5 & 56.9 & 54.4 & 54.4 & 53.0 & 56.5 & 55.0 & 51.9 & 54.9 & 54.9 & 13.6 & 9.1 & 17.3 &  16.0 \\
				\cline{2-17}%\hline
				& \changes{Median} & \changes{58.0} & \changes{60.7} & \changes{67.5} & \changes{62.9} & \changes{62.9} & \changes{60.6} & \changes{66.0} & \changes{64.5} & \changes{57.9} & \changes{62.6} & \changes{62.7} & \changes{5.8} & \changes{1.4} & \changes{9.6} & \changes{9.4} \\
				\cline{2-17}%\hline
				& \changes{Std. Dev.} & \changes{33.5} & \changes{34.0} & \changes{33.8} & \changes{34.2} & \changes{34.2} & \changes{34.4} & \changes{34.4} & \changes{34.1} & \changes{33.2} & \changes{34.1} & \changes{34.3} & \changes{17.9} & \changes{14.0} & \changes{19.8} & \changes{18.2} \\
				\hline
				\hline
				\multirow{3}{*}{\rotatebox[origin=c]{90}{\#Calls}}
				&Avg. & 667.8 & 676.3 & 641.2 & 651.2 & 651.2 & 660.1 & 612.1 & 649.1 & 720.9 & 704.5 & 667.1 & 872.6 & 979.7 & 931.6 & 913.5  \\
				\cline{2-17}%\hline
				&\changes{Median}  & \changes{84} & \changes{77} & \changes{64} & \changes{73} & \changes{73} & \changes{62} & \changes{60} & \changes{62} & \changes{70} & \changes{64} & \changes{71} & \changes{168} & \changes{150} & \changes{143} & \changes{158} \\
				\cline{2-17}%\hline
				&\changes{Std. Dev.} & \changes{1734.1} & \changes{1794.4} & \changes{1960.6} & \changes{1676.5} & \changes{1676.5} & \changes{1757.4} & \changes{1632.3} & \changes{1755.1} & \changes{2131.2} & \changes{2140.8} & \changes{2053.0} & \changes{1889.4} & \changes{2062.0} & \changes{2141.8} & \changes{2046.7}\\
				\hline
	\end{tabular}
    \caption{Fraction of overall execution time (in $\%$), reduction ratio (in $\%$)\changes{, and the total number of generalization attempts. Average, Median, and Standard Deviation.} Original PDR.}
	\label{tab:genratioorig}
\end{table*}

%\vspace{-2mm}
\paragraph{Full PDR runs, execution times, PO reduction ratios\changes{, and number of PO generalization attempts}}\label{para:full_pdr_ratios}

\changescs{
In Table~\ref{tab:genratioorig} we present the fraction of the overall execution time that the methods considered in Fig.~\ref{fig:gen_categories} required during \emph{full} PDR runs
(\ie{} the time required for generalization divided by the overall execution time).
Moreover, we give
information on their
reduction ratios\changes{, and the number of generalization attempts} during full PDR runs.}
\changes{We report on average and median values, as well as the standard deviation.}
For computing the reduction ratios \changes{and generalization attempts} we consider only benchmarks solved by all those methods.
\changescs{It is interesting to correlate the data in Table~\ref{tab:genratioorig} with the overall performance data
in Fig.~\ref{fig:gen_categories}. For instance, \emph{ILP cover} and \emph{greedy cover} need a large fraction of the overall runtime without providing high reduction rates which explains their poor overall performance.
Both lifting and \emph{IGBG} are very runtime efficient (their \changes{average} fractions of execution times are 2.2 \% resp.~2.0 \%)
with good reduction rates, whereas
the advantage of \emph{IGBG} can be explained by its higher PO reduction \changes{ratios}.
Enhancing lifting with restricted (`TIP-like') literal rotation leads to a moderate increase in
the fraction of runtime for PO generalization, but pays-off by an increased reduction rate
(and an improved overall performance as well).
}
\emph{MS01X} shows the best \changes{average} reduction rate (56.5 \%), but has high \changes{average} execution times (60.2 \%).
For this reason \emph{MS01X} does not have the best overall performance as can be seen in Fig.~\ref{fig:gen_categories}.
Using the \emph{MS01X} / \emph{IGBG} heuristics successfully reduces the \changes{average} fraction of execution time
to 20.5 \% while keeping a high \changes{average} reduction ratio of 55.0 \%. By this
the \emph{MS01X} / \emph{IGBG} heuristics ends up with the best overall performance.
\changes{In general, lower reduction ratios seem to significantly increase the total number of required \po{} generalization attempts -- as can be observed for all cover approaches in particular.
Overall efficiency, however, very much depends on how costly these \po{} generalizations are in terms of runtime.
An example would be MS01X which achieves the lowest number of \po{} generalization calls (median and average) due to its strong reduction ratio -- since these calls are very costly though, it is still outperformed by more cost-efficient techniques.
The large standard deviations as well as the differences between average and median values (especially for the number of \po{} generalization calls)
show that the HWMCC benchmark set is pretty diverse. It apparently contains a large fraction of benchmarks which are easy for PDR, but
contains difficult instances as well.
}

Note that -- compared to the study with single POGPs -- the \changes{average} reduction ratios arrive at much higher values.
This can be explained by the fact that for single POGPs we only considered instances
which can also be solved by the MaxQBF approach.
High reduction ratios result in a high number of satisfied MaxQBF soft clauses.
A growing number of such clauses is quite challenging for the incorporated experimental MaxQBF solver
leading to timeouts on many HWMCC benchmarks.

\begin{table}[t]
	\centering
\scriptsize
	\begin{tabular}{|c|c|c|c|c|c|}
				\multicolumn{2}{c|}{} & Struct. & GreedyQBF & MaxQBF$^{\txt{free}}$ & MaxQBF$^{\txt{fix}}$ \\
				\hline
				\multicolumn{6}{c}{}\\[-7pt]
				\hline
				\multirow{3}{*}{\rotatebox[origin=c]{90}{\parbox{8mm}{$\%$ Exe. \\Time}}}
				&Avg. & 16.4 & 96.3 & 73.2 & 67.8 \\
				\cline{2-6}%\hline
				&\changes{Median} & \changes{1.8} & \changes{94.7} & \changes{84.5} & \changes{69.4} \\
				\cline{2-6}%\hline
				&\changes{Std. Dev.} & \changes{8.7} & \changes{16.0} & \changes{28.3} & \changes{33.4} \\
				\hline
				\hline
				\multirow{3}{*}{\rotatebox[origin=c]{90}{\parbox{8mm}{$\%$ Red. \\Ratio}}}
                &Avg. &  8.6 & 12.8 & 11.4 & 12.8 \\
				\cline{2-6}%\hline
                &\changes{Median} &  \changes{0.5} & \changes{3.2} & \changes{3.2} & \changes{3.2} \\
				\cline{2-6}%\hline
               & \changes{Std. Dev.} &  \changes{8.0} & \changes{10.6} & \changes{9.1} & \changes{10.5} \\
				\hline
				\hline
				\multirow{3}{*}{\rotatebox[origin=c]{90}{\#Calls}}
				&Avg. &  1481.1 & 1064.0 & 1066.0 & 1057.8 \\
				\cline{2-6}%\hline
				&\changes{Median} & \changes{35}  & \changes{25} & \changes{25} & \changes{25} \\
				\cline{2-6}%\hline
				&\changes{Std. Dev.}&  \changes{4996.6} & \changes{3341.2} & \changes{3203.7} & \changes{3302.2} \\
				\hline
	%\hline
	\end{tabular}
	%\vspace{1mm}
    \caption{Fraction of overall execution time (in $\%$), reduction ratio (in $\%$)\changes{, and the total number of generalization attempts. Average, Median, and Standard Deviation.} \emph{Reverse PDR}.}
	\label{tab:genratiorev}
\end{table}

\subsubsection{Reverse PDR}\label{ssse:revpdr}
We recall from Sect.~\ref{sec:specrev} that generalization of \pos{} in Reverse PDR is only possible with both the QBF-based methods and the structural method from~\cite{SeSc:2018}.

\paragraph{Full Reverse PDR runs, overall performance}

\begin{figure}
	\begin{center}
	\includegraphics[width=4.5cm]{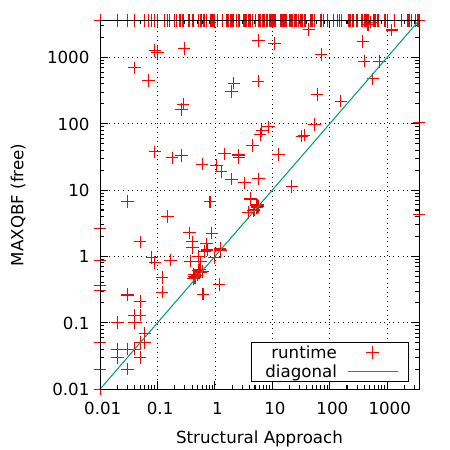}\\
	\end{center}
	\caption{MAXQBF$^{\txt{free}}$ / Structural \changes{\cite{SeSc:2018}}}
\label{fig:1517maxqbfvsrevpostruct}
\end{figure}
In the case of Reverse PDR, the structural generalization method outperforms both \emph{MaxQBF} and \emph{greedy QBF}
in terms of
\changescs{solved instances.}
The structural method even completely dominates \emph{greedy QBF}.
This can be explained by the large \changescs{computational} effort invested by the stronger generalization methods.
Though
there exist some benchmarks where \emph{MaxQBF$^{free}$} performs better than the structural approach.
\changescs{Two of them cannot even be solved within the timeout using the structural approach.}
For a more detailed comparison of execution times, we refer to Fig.~\ref{fig:1517maxqbfvsrevpostruct}.

%\vspace{-2mm}
\paragraph{Full Reverse PDR runs, execution times and PO reduction ratios}
\changescs{Table~\ref{tab:genratiorev} shows the average fractions of the overall execution times used by the respective generalization approaches
as well as their reduction ratios. The structural approach has the smallest reduction ratios, but (as expected)
it requires much less runtime than the QBF / MaxQBF based methods.}
We note that reduction ratios with Reverse PDR are \changescs{in general} significantly smaller than with original PDR.
This can be explained by the limitations of left-unique transition relations (see Sect.~\ref{sec:specrev}).

\subsubsection{Invariant Constraints}\label{ssse:invariant_constraints}

\begin{figure}[tb]%{wrapfigure}[11]{r}{9.0cm} %[number of narrow lines] {placement} [overhang] {width of figure}
	\begin{center}
    \vspace{-5mm}
	%\hspace*{-5mm}
	\includegraphics[width=8.5cm]{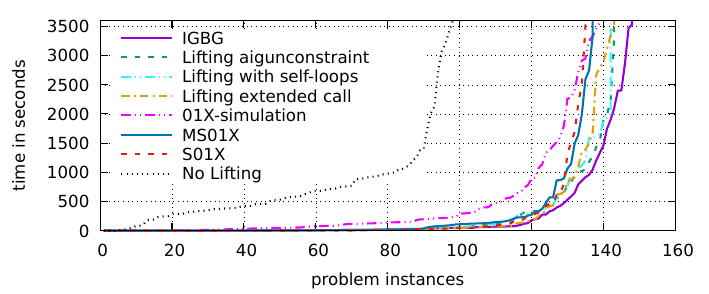}\\
	\end{center}
	\vspace{-3mm}
	%\hspace*{-5mm}
	\caption{HWMCC'19, with invariant constraints.}
\label{fig:cactus19}
\end{figure}%{wrapfigure}
For standard \emph{ic3ref} and HWMCC'19 benchmarks with invariant constraints
we observed incorrect results\footnote{\changes{For instance: ic3ref reports Unsafe instead of Safe on wolf/2019C/qspiflash qflexpress divfive-
p072.aig or wolf/2019C/qspiflash qflexpress divfive-p077.aig from HWMCC'19.}}. This can be explained by the observations on lifting
made in Sect.~\ref{sec:selfloops}. So we had to deactivate lifting for \emph{ic3ref} to get correct results.
In Fig.~\ref{fig:cactus19} we compare the execution times of the three \emph{lifting} variants from Sect.~\ref{sec:selfloops} 
against the admissible 01X-techniques and \emph{IGBG} as well as to \emph{ic3ref} with deactivated lifting.
By `Lifting extended call' we denote the extended query $\txt{SAT?}[m \wedge i \wedge \hat{T} \wedge (\neg C \vee \neg d')]$ without `repairing' the transition relation.
The lifting variants greatly outperform standard \emph{ic3ref} with deactivated lifting. Furthermore, the lifting variants apparently come with only minimal overhead and similar execution times, thus \emph{lifting} outperforms \emph{01X-simulation}, \emph{S01X} and  \emph{MS01X}. \emph{IGBG} does not need any changes and
achieves the best performance.

\subsection{AI Planning}\label{sse:ai_planning}
We adjusted an existing PDR implementation called \emph{minireachIC3}~\cite{MIC3:2014} which has been used for planning tasks~\cite{Suda:2014}. We used 1641 STRIPS benchmarks from past IPC events (from 1998 to 2011). To transform the STRIPS benchmarks into the input format (DIMSPEC)
of \emph{minireachIC3} we used the $\exists$-step parallel encoding scheme of the SAT-based planner \emph{Mp}~\cite{Rint:2012} as found best for \emph{minireachIC3}~\cite{Suda:2014}. Since the previous evaluation on IPC benchmarks from~\cite{Suda:2014} came to the conclusion that Reverse PDR
(for general transition relations, the reverted direction does not imply left-uniqueness)
is the favorable configuration of \emph{minireachIC3}, we also use this configuration.

We compared standard \emph{minireachIC3} \changescs{(which does not include PO generalization)} with
the three cover approaches, \emph{GeNTR}, \emph{MaxQBF} as well as \emph{greedy QBF}.
We observed that in AI Planning the QBF-based approaches are much more competitive than
in Hardware Model Checking.
However,
many of the planning benchmarks seem to have
only little potential for generalization of \pos{}. The best average reduction ratio -- achieved by \emph{MaxQBF} -- on 695 of 1641 solved IPC benchmarks
is $4.01\,\%$.

\begin{table}
	\centering
\scriptsize
	\begin{tabular}{c|c|c|c|c|c|}
		Technique & Solved & SAT & UNSAT & Timeout & Memout \\
		\hline
		\hline
		SAT Cover & 948 & 934 & 14 & 638 & 55 \\
		\hline
		Greedy Cover & 940 & 926 & 14 & 646 & 55 \\
		\hline \hline
		{\bf Standard} & 939 & 925 & 14 & 643 & 59 \\
		\hline \hline
		GeNTR & 913 & 899 & 14 & 614 & 114 \\
		\hline
		ILP Cover & 900 & 886 & 14 & 722 & 19 \\
		\hline
		MaxQBF & 695 & 681 & 14 & 927 & 19 \\
		\hline
		GreedyQBF & 646 & 636 & 10 & 968 & 27 \\
		\hline
		%\hline
	\end{tabular}
    \caption{\emph{minireachIC3} on IPC Benchmarks.}
	\label{tab:ipcresults}
\end{table}

\changescs{Nevertheless, we present the overall performance of the different \po{} generalization techniques
for the complete set of benchmarks in Table~\ref{tab:ipcresults}.
Since many of the planning benchmarks have
only little potential for generalization of \pos{} and show only small average reduction ratios,
expensive methods like MaxQBF result in high cost for many benchmarks, but do not help much.
However, SAT-based cover, which is much less expensive, is able to consistently improve on standard
\emph{minireachIC3}. It solves 948 instances overall (compared to 939 instances with standard
\emph{minireachIC3}).
On the other hand, compared to SAT-based cover, the MaxQBF approach is able to solve 21 unique instances in total and additionally achieves better execution times on a number of benchmarks.}

\changescs{As presented in Table~\ref{tab:selection} there are planning domains} -- that allow significant reduction of \pos{} -- for which the generalization approaches greatly improve the performance of \emph{minireachIC3}:
Among the cover approaches and \emph{GeNTR} (which is implicitly a cover approach as well) SAT-based cover is
the best and clearly outperforms standard \emph{minireachIC3}.
Furthermore, \changescs{for those planning domains} trading computation time against generalization capabilities pays off and MaxQBF
performs even better than the cover approaches on three domains.
It
dominates standard \emph{minireachIC3} on 2002-DEPOTS, 2006-PIPESWORLD, 2011-BARMAN, and 2011-FLOORTILE.
Numbers in bold face indicate the best performance for the respective domain.

\begin{table}[t]
	\centering
   \scriptsize
   \setlength{\tabcolsep}{1.2mm} % Spaltenabstand anpassen
	\begin{tabular}{|c|m{3.2mm}|c|c|c|c|c|c|c|}
				Domain & Inst. & \shortstack{Stan- \\ dard} & \shortstack{Max \\ QBF} & \shortstack{Greedy \\ QBF} & \shortstack{Greedy \\ Cover} & \shortstack{Ge- \\ NTR} & \shortstack{ILP\\Cover} & \shortstack{SAT \\ Cover} \\
				\hline
				\hline
				2002-DEPOTS & 20 & 4 & \textbf{8 (4)} & \textbf{6 (2)} & 4 (0) & 4 (0) & 4 (0) & 4 (1)\\
				\hline
				2006-PIPESW. & 45 & 6 & 6 (1) & 3 & 6 (0) & \textbf{7 (1)} & 4 (0) & 6 (1)\\
				\hline
				2008-SCANALY. & 30 & 22 & 16 (0) & 15 (0) & 22 (2) & \textbf{23 (2)} & \textbf{23 (1)} & \textbf{23 (3)}\\
				\hline
				2011-BARMAN & 20 & 12 & \textbf{13 (2)} & 0 & \textbf{14 (3)} & 12 (0) & 7 (2) & 10 (6)\\
				\hline
				2011-FLOORT. & 20 & 17 & \textbf{20 (3)} & 15 (1) & 16 (1) & 17 (0) & 18 (2) & 19 (2)\\
				\hline
				2011-SCANALY. & 20 & 14 & 9 (0) & 6 (0) & 12 (0) & 13 (0) & \textbf{15 (1)} & \textbf{15 (3)}\\
				\hline
	%\hline
	\end{tabular}
	\vspace{1mm}
	\caption{Solved instances of generalization techniques on selected IPC domains. The number of uniquely solved instances (\wrt{} Standard) is written in parentheses.}
    \vspace{-5mm}
	\label{tab:selection}
\end{table}

\changes{For a detailed analysis we chose two examples where MaxQBF in \emph{minireachIC3} had a strong effect and analyze further why this is the case.}

Firstly, we analyze the problem \emph{depotprob4398} of the IPC domain \emph{DEPOTS-2002}. This could be solved within $51.33\,s$ by using \emph{MaxQBF} for \po{} generalization, finding a plan of length $13$ (longer than the trace, due to \po{} forward pushing), whereas standard \emph{minireachIC3} ran into timeout ($3,600\,s$). With \emph{MaxQBF}, only $6$ time frames were opened, while the standard version \changescs{ran into timeout when it was still working on time frame number $9$}. The version using \emph{MaxQBF} had to learn only $1,226$ clauses with an average size of $17.2$, whereas the standard version had $86,932$ learned clauses with an average size of $23.0$ after $3,600\,s$.

Secondly, we consider the problem \emph{instance-1} of the IPC domain \emph{BARMAN-2011}. With \emph{MaxQBF} we were able to find a plan of length $142$ after $82.37\,s$, again standard \emph{minireachIC3} ran into timeout ($3,600\,s$).
With \emph{MaxQBF}, \emph{minireachIC3} advanced to time frame $15$, without \emph{MaxQBF}, it had $13$ open time frames
\changescs{within}
$3,600\,s$. The \emph{MaxQBF} version required $1,289$ learned clauses with an average size of $19.1$ literals, while the standard version reached
\changescs{the timeout}
with $104,503$ learned clauses
\changescs{having}
an average size of $20.2$ ($103,400$ of them were learned in only two time frames $5$ and $6$).

\section{Conclusions and Future Work}
\label{sec:conclusions}
We presented a comprehensive study on the generalization of \pos{} in PDR.
We discussed the complexity of the problem as well as limitations and applicability
of various techniques on different domains.
It turned out that techniques which have not been used in PDR so far as well as new and more exact techniques
based on MaxSAT are able to beat well-known standard techniques for the generalization of \po{}s
like lifting and 01X-simulation. We expect to be able to improve the obtained results even more
using further improvements of solver technology like incremental reuse of cardinality constraints in MaxSAT
solvers. An exact solution for general transition relations could be provided using a reduction to MaxQBF.
Research on MaxQBF solving is still in its infancy and we had to use a rather immature solver
for our experiments.
We hope that with new applications we can stimulate further research in this direction.
We successfully explored the cooperation between different generalization approaches with a combination of
\emph{MS01X} and \emph{IGBG}. We believe that there is room for further improvements by combining
various techniques.
\changescs{One option is to}
dynamically switch between different methods based on their success, 
\changescs{another option is to }
perform first generalizations with weaker and less expensive methods, followed by stronger methods building on the results of the former.
Moreover, we assume that our results can be useful for other domains requiring
SMT with theory reasoning.

\bibliographystyle{IEEEtran}
\bibliography{literature}
\end{document}